\documentclass[12pt]{iopart}

\usepackage[a4paper, margin=0.8in]{geometry}
\usepackage{iopams}  
\usepackage{color}
\usepackage[dvipsnames]{xcolor}
\usepackage{mathrsfs, amssymb, amsthm}
\usepackage{graphicx}
\usepackage[colorlinks=true, linkcolor=NavyBlue, citecolor=MidnightBlue, urlcolor=CornflowerBlue]{hyperref}
\providecommand{\eprint}[2][]{\href{https://arxiv.org/abs/#2}{\path{arXiv:#2}}}

\usepackage{etoolbox}
\usepackage{overpic}
\usepackage{soul}
\usepackage{booktabs}
\usepackage{tabularx}

\makeatletter
\def\@mkboth#1#2{}
\newlength\appendixwidth
\preto\appendix{\addtocontents{toc}{\protect\patchl@section}}
\newcommand{\patchl@section}{%
  \settowidth{\appendixwidth}{\textbf{Appendix }}%
  \addtolength{\appendixwidth}{1.5em}%
  \patchcmd{\l@section}{1.5em}{\appendixwidth}{}{}%
  \patchcmd{\l@subsection}{2.3em}{\appendixwidth}{}{}%
}
\makeatother

\newcommand{\scri}{\mathscr{I}}
\newcommand{\sech}{\mathrm{sech}}
\providecommand{\eqref}[1]{(\ref{#1})}

\begin{document}

\title[Hyperboloidal evolution for scalar scattering in Minkowski space]{Hyperboloidal evolution for scalar scattering in Minkowski space}

\author{Ekrem Demirbo\u{g}a, An\i l Zengino\u{g}lu}
\address{Institute for Physical Science and Technology, University of Maryland, College Park, MD 20742, USA}
\ead{ekrem@umd.edu, anil@umd.edu}

\begin{abstract}
We develop a time-domain numerical framework for global scalar wave scattering in Minkowski spacetime. The main contribution is an exact conformal matching of three compactified regions: a past hyperboloidal domain attached to $\mathscr I^-$, a Penrose domain covering a neighborhood of spatial infinity $i^0$, and a future hyperboloidal domain attached to $\mathscr I^+$. The matching surfaces are identical conformal hypersurfaces in the adjacent charts. This yields a global evolution scheme connecting $\mathscr I^-$, the neighborhood of $i^0$, and $\mathscr I^+$ without artificial timelike outer boundaries and without interpolation between scri-fixing gauges.

We implement the construction for spherically symmetric scalar waves, including free propagation, localized linear scattering potentials such as the P\"oschl--Teller potential, and semilinear wave equations with cubic, quintic, and septic nonlinearities. The numerical experiments demonstrate stable propagation across the matching interfaces, direct extraction of radiation at $\mathscr I^+$, and fourth-order convergence for the free and linear-potential tests. The quintic and septic nonlinear tests exhibit approximately fourth-order convergence and recover the expected late-time tail rates. The cubic case, by contrast, shows only first-order convergence, revealing a limitation of our treatment near compactified boundaries when the conformally rescaled nonlinear source remains non-vanishing. These results validate the conformal matching strategy for long-time simulations, while identifying the boundary regularity issues that must be addressed using a more robust treatment of spatial infinity.
\end{abstract}

\vspace{2pc}

%
%
%

\tableofcontents

\section{Introduction}
\label{sec:introduction}

Scattering in asymptotically flat spacetimes is most naturally formulated as a problem at null infinity to determine the map between asymptotic data \cite{penrose_asymptotic_1963, penrose_zero_1965, friedlander_radiation_1980, friedrich1981asymptotic, baez_global_1990,mason_nicolas_2004}. Incoming radiation is prescribed at past null infinity $\mathscr I^-$ and outgoing radiation is read off at future null infinity $\mathscr I^+$. In this global formulation the radiation fields at infinity, rather than finite-radius waveforms or finite-time Cauchy data, are the primary objects. The corresponding scattering operator is schematically the map $\mathcal S: \mathcal D_-(\mathscr I^-) \longrightarrow \mathcal D_+(\mathscr I^+)$, between suitable spaces of incoming and outgoing radiation fields.

Standard numerical evolutions begin with data on a finite spacelike hypersurface and extract radiation at a finite coordinate radius, followed by extrapolation, Cauchy-characteristic extraction, or similar post-processing methods  \cite{bishop_rezzolla_2016,bishop_gomez_lehner_maharaj_winicour_1996, moxon2023spectre, bernuzzi2025perturbative}. However, this approach does not provide direct access to the intrinsic scattering map at the conformal boundary. Finite-radius extraction introduces gauge and extrapolation ambiguities, while finite-time Cauchy data can contain spurious components associated with a mismatch between the prescribed initial data and the desired asymptotic structure of the spacetime \cite{Lovelace_2009,Pfeiffer_2003,compere_asymptotic_2023,valiente2024d}. It is therefore desirable to construct a formulation in which null infinity is part of the computational domain rather than an external limiting surface.

Hyperboloidal compactification gives direct numerical access to null infinity by using spacelike hypersurfaces that asymptote to $\mathscr I$ and by compactifying the outgoing direction \cite{friedrich1983cauchy, frauendiener2004conformal}. Radiation extraction at null infinity becomes numerical evaluation at a finite grid shell. However, a global scattering calculation requires not only access to a single component of null infinity but a past foliation adapted to $\mathscr I^-$ and a future foliation adapted to $\mathscr I^+$. The transition between these domains requires control over the neighborhood of spatial infinity $i^0$.

The main contribution of this paper is a geometric and numerical construction of such a global scattering problem in flat spacetime. We use three fixed conformal domains: a past hyperboloidal domain $\mathcal H_-$ attached to $\mathscr I^-$, a central Penrose domain $\mathcal P$ covering a neighborhood of $i^0$, and a future hyperboloidal domain $\mathcal H_+$ attached to $\mathscr I^+$. The matching is exact at the level of the conformal geometry in the sense that the transition hypersurfaces between the three domains agree. The resulting evolution covers the full scattering problem, avoids artificial timelike outer boundaries, and does not require interpolation between the domains.

The global scattering problem in Minkowski spacetime has gained significant interest in the mathematical literature \cite{lindblad_schlue_2023,lindblad_schlue_2025, Kadar:2025xmo, marajh_taujanskas_valientekroon_2025} but there are very few numerical implementations for this problem \cite{zenginouglu2026penrose, duarte2026numerical}. The closest time-domain implementation related to our work is the numerical framework of Duarte-Baptista, Va{\~n}{\'o}-Vi{\~n}uales, and del R{\'i}o for particle creation in accelerating toy models \cite{duarte2026numerical}. Their method uses two hyperboloidal domains to propagate scalar modes from $\mathscr I^-$ to $\mathscr I^+$, demonstrating that scattering data for particle creation can be computed directly at null infinity. In their implementation, past and future hyperboloidal foliations are connected through interpolation between the two descriptions and the neighborhood of spatial infinity is not part of the computational domain. The construction developed here addresses these issues by inserting a Penrose domain so that the past and future hyperboloidal evolutions are connected by exact matching rather than by an interpolation.

Our work is also motivated by the fully nonlinear numerical treatment of gravitational-wave scattering using Friedrich's generalized conformal field equations \cite{frauendiener2025fully, friedrich1998gravitational}.  In their method, the computational domain includes both $\mathscr I^-$ and $\mathscr I^+$, demonstrating that nonlinear gravitational scattering can be treated numerically in a conformal framework.  However, incoming radiation is specified indirectly through a timelike boundary in the unphysical spacetime, and the authors note that a fully global scattering problem would require the inclusion of spacelike infinity. Our work focuses on the flat spacetime model to demonstrate long-time scattering simulations by using hyperboloidal scri-fixing combined with a propagation through a central conformal region that covers a neighbourhood of spatial infinity. The long-time simulations are made possible by using time-translation invariant hyperboloidal domains fixing scri on the numerical grid \cite{zenginoglu_hyperboloidal_2008}. This domain-decomposition strategy suggests a possible route toward future Einstein evolutions once the hyperboloidal evolution problem is under control \cite{Zenginoglu:2008pw, Vano-Vinuales:2014koa, gasperin_gautam_hilditch_vano_2020, gautam_vano_hilditch_bose_2021, peterson_gautam_vano_hilditch_2024, Vano-Vinuales:2024tat, Alvares:2025pbi}.

The paper is organized as follows. In Section~\ref{sec:geometry}, we present the conformal geometry and the coordinate construction underlying the three computational domains. In Section~\ref{sec:conformal-semilinear}, we introduce the model problems used to test the framework, which include the free conformal wave equation, localized linear potentials, and semilinear cubic, quintic, and septic equations. In Section~\ref{sec:numerics}, we describe the numerical implementation and the matching procedure. In Section~\ref{sec:results}, we present the numerical results. Finally, in Section~\ref{sec:discussion}, we summarize the construction and outline possible extensions to more general spacetimes and field equations.

\section{Geometric Framework}\label{sec:geometry}

\subsection{Infinity and Conformal Compactification}

In the study of scattering phenomena in asymptotically flat spacetimes, the idealized notion of infinity plays a central role. In Minkowski space, the causal structure can be compactified into a finite diagram using a conformal rescaling of the metric~\cite{penrose_asymptotic_1963}. Under this transformation, spatial infinity $i^0$, past null infinity $\mathscr{I}^-$, and future null infinity $\mathscr{I}^+$ are brought to finite coordinate locations, which make conformal methods appealing for numerical computations.

Penrose diagrams provide a compact representation of spacetime, where null rays travel at $45^\circ$ angles and infinity becomes part of the boundary. Incoming future-directed null rays enter the spacetime through $\mathscr{I}^-$; outgoing future-directed null rays terminate at $\mathscr{I}^+$. Spatial infinity $i^0$ marks the limit of spacelike geodesics at infinite distance. Accurately capturing the dynamics near these asymptotic regions is essential for a faithful description of global scattering problems.

\subsection{Penrose coordinates}\label{sec:standardPenrose}
\begin{figure}[ht]
\centering
\includegraphics[width=0.3\textwidth]{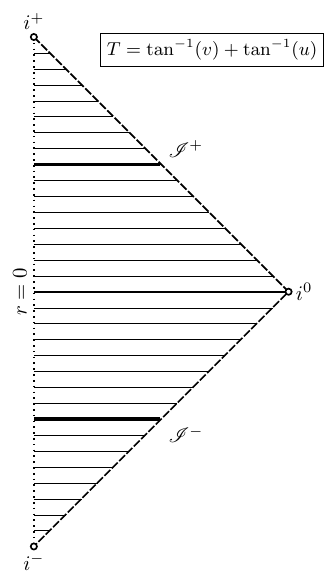}\hfill
\includegraphics[width=0.3\textwidth]{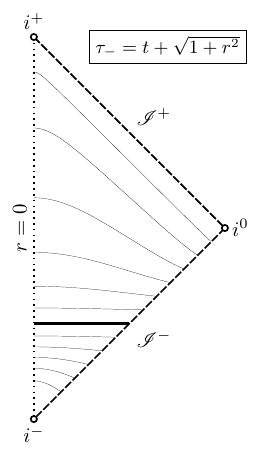}\hfill
\includegraphics[width=0.3\textwidth]{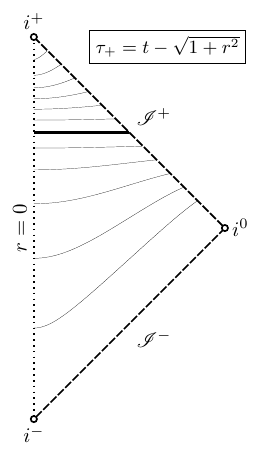}
\    \caption{Coordinate systems used in the numerical construction: 
    (a) Penrose coordinates,
    (b) past hyperboloidal coordinates, 
    (c) future hyperboloidal coordinates. The hypersurfaces $T=\pm \pi/2$ in the Penrose diagram coincide with the hyperboloidal slices $\tau_\pm=0$ in the respective hyperboloidal diagrams (emphasized as thick, horizontal black lines).
    }
    \label{fig:coordinates}
\end{figure}

We review the Penrose compactification of Minkowski spacetime \cite{penrose_asymptotic_1963, Kroon_2023, carroll2019spacetime, Zenginoglu:2024bzs}. The Minkowski metric in spherical coordinates is written as,
\begin{equation}
\eta = -dt^2 + dr^2 + r^2\, d\sigma^2,
\end{equation}
where $d\sigma^2$ denotes the round metric on the unit two-sphere. Introducing null coordinates
\begin{equation}
  u = t-r, \qquad v = t+r,
\end{equation}
the metric takes the form
\begin{equation}
  \eta = -\,du\,dv + \frac{(v-u)^2}{4}\, d\sigma^2.
\end{equation}
Following the standard Penrose compactification, we map the null directions to finite ranges via
\begin{equation}
  u=\tan U,\qquad v=\tan V,
  \qquad U,V\in\left(-\frac{\pi}{2},\frac{\pi}{2}\right),
\end{equation}
and define compactified time and radius coordinates
\begin{equation}
  T = U+V,\qquad R = V-U
  \quad\Longleftrightarrow\quad
  U=\frac{T-R}{2},V=\frac{T+R}{2}.
\end{equation}
Note that the level sets of $T$ are hyperboloidal for $T\in (-\pi, \pi) \setminus \{0\}$ \cite{Zenginoglu:2024bzs}. This property is essential in our construction, and can also be clearly seen from the first diagram in Figure~\ref{fig:coordinates}.

Minkowski spacetime is conformal to a region of the Einstein static universe with line element
\begin{equation}
g := \Omega_P^2 \eta = -dT^2 + dR^2 + \sin^2 R\, d\sigma^2,
\label{eq:penrose_metric}
\end{equation}
and conformal factor $\Omega_P = 2\cos U \cos V = \cos T + \cos R$.
The physical region $r\ge 0$ corresponds to $R\ge 0$, and the coordinate ranges imply the standard diamond $0\le R<\pi,\qquad |T|+R<\pi$. Null infinity is the conformal boundary $\Omega_P=0$, i.e.\ $\cos T+\cos R=0$, which splits into $\mathscr{I}^+: T+R=\pi$ and $\mathscr{I}^-: T-R=-\pi$.
In particular, on the slices $T=\pm \pi/2$ one has $\cos T=0$, so $\Omega_P=0$ occurs at $R=\pi/2$; i.e.~$T=\pm\pi/2$ intersects $\mathscr{I}^\pm$ at $R=\pi/2$. These are the slices that we employ to match to hyperboloidal evolution described next.


\subsection{Stationary Hyperboloidal Coordinates}

Hyperboloidal foliations have been effective in evolving propagating fields in asymptotically flat spacetimes \cite{zenginoglu_tails_2008, bizon2010saddle, ZENGINOGLU20112286,ripley2021numerical, Panosso_Macedo_2024, bourg2025quadratic}. Instead of using standard Cauchy slices that extend to spatial infinity, hyperboloidal slices are tilted toward null infinity, reaching $\mathscr{I}^\pm$. Combining two families of hyperboloidal foliations, we can prescribe incident boundary data at $\mathscr{I}^-$ and the extraction of radiation at $\mathscr{I}^+$ without imposing artificial boundary conditions. While this property is already satisfied by Penrose coordinates, the time-dependent nature of Penrose coordinates make long-time simulations cumbersome.

Instead, we switch to time-translation invariant (stationary) hyperboloidal foliations \cite{zenginoglu_hyperboloidal_2008}. We define two distinct hyperboloidal time coordinates: a past hyperboloidal time $\tau_-$ adapted to $\mathscr{I}^-$ and a future hyperboloidal time $\tau_+$ adapted to $\mathscr{I}^+$. The requirement of stationarity implies that a single foliation cannot cover both asymptotic regions to the past and the future due to the vanishing of the timelike Killing field at spatial infinity \cite{Zenginoglu:2024bzs}. We describe a solution to this problem in Sec.~\ref{sec:penrose-hyperboloidal-matching}.

We introduce a compactification of the radial coordinate via $r(\rho) = \tan \rho$ with $\rho \in [0, \pi/2)$. Infinity is mapped to a finite coordinate location $\rho = \pi/2$ on the time slices $\tau_\pm$ defined through the hyperboloidal coordinate transformations
\begin{equation}\label{eq:past_hyperboloidal}
t = \tau_- - \sqrt{1 + r(\rho)^2} = \tau_- - \frac{1}{\cos\rho}, \qquad \rm{(past\ hyperboloidal)},
\end{equation}
\begin{equation}\label{eq:future_hyperboloidal}
t = \tau_+ + \sqrt{1 + r(\rho)^2}= \tau_+ + \frac{1}{\cos\rho}, \qquad \rm{(future\ hyperboloidal)}.
\end{equation}
The metric takes a conformally rescaled form
\begin{equation}\label{eq:hyperboloidal_metric}
g = \Omega_H^2 \eta = -\Omega_H^2 \, \rm{d}\tau_\pm^2 \mp 2 \sin \rho \, \rm{d}\tau_\pm \rm{d}\rho + \rm{d}\rho^2 + \sin^2 \rho \, \rm{d}\sigma^2,
\end{equation}
where $d\sigma^2$ is the metric on the unit sphere and the conformal factor is $\Omega_H = \cos \rho$. For later reference, we note the explicit expressions for the null coordinates in these charts. For future hyperboloidal foliations, we have
\begin{equation}\label{eq:uv_future}
u = \tau_+ + \frac{\cos\rho}{1+\sin\rho}, \qquad
v = \tau_+ + \frac{1+\sin\rho}{\cos\rho},
\end{equation}
For past hyperboloidal foliations, we have
\begin{equation}\label{eq:uv_past}
u = \tau_- - \frac{1+\sin\rho}{\cos\rho}, \qquad
v = \tau_- - \frac{\cos\rho}{1+\sin\rho},
\end{equation}
Inspecting the properties of the characteristics at null infinity, $\rho=\pi/2$, we see that for future hyperboloidal foliations the outgoing characteristic is regular while the incoming characteristic blows up and vice versa for the past hyperboloidal foliation. Therefore, incoming data can only be prescribed in a past hyperboloidal foliation.

\subsection{Matching between hyperboloidal and Penrose coordinates}
\label{sec:penrose-hyperboloidal-matching}

The key observation for a global evolution in Minkowski spacetime is that the Penrose slices $T=\pm\pi/2$ coincide with the hyperboloidal slices $\tau_\pm=0$. The inverse Penrose transformation is
\begin{equation}
  t=\frac{\sin T}{\cos T+\cos R},
  \qquad
  r=\frac{\sin R}{\cos T+\cos R}.
  \label{eq:inverse-penrose}
\end{equation}
On the slices $T=\pm\pi/2$, this reduces to
\begin{equation}
  t=\pm\frac{1}{\cos R},
  \qquad
  r=\tan R.
  \label{eq:Tpm-slice}
\end{equation}
Therefore, these Penrose slices satisfy $t^2-r^2=1$. Comparing with
\begin{equation}
  t=\tau_- -\sqrt{1+r^2},
  \qquad
  t=\tau_+ +\sqrt{1+r^2},
\end{equation}
we see that $T=-\pi/2$ is the same hypersurface as $\tau_-=0$ and $T=\pi/2$ is the same hypersurface as $\tau_+=0$. Since both charts use the same compactifying radial coordinate on the matching surface, $r=\tan R=\tan\rho$, the matching is pointwise:
\begin{equation}
  R=\rho,
  \qquad
  (t,r)=(\pm\sec R,\tan R),
  \qquad
  \tau_\pm=0.
  \label{eq:rho-equals-R-on-matching}
\end{equation}

The resulting spacetime decomposition consists of three conformal evolution domains,
\begin{eqnarray*}
  \mathcal H_- &:=& \{(\tau_-,\rho): \tau_-\le 0,\ 0\le\rho\le\pi/2\}, \\
  \mathcal P &:=& \{(T,R): -\pi/2\le T\le\pi/2,\ 0\le R\le R_{\rm b}(T)\},\\
  \mathcal H_+ &:=& \{(\tau_+,\rho): \tau_+\ge 0,\ 0\le\rho\le\pi/2\},
\end{eqnarray*}
 where $R_{\rm b}(T)=\pi-|T|$. The past and future hyperboloidal domains $\mathcal H_-$ and $\mathcal H_+$ are scri-fixed: their outer boundaries are located at $\rho=\pi/2$, representing $\mathscr I^-$ and $\mathscr I^+$, respectively. The Penrose domain $\mathcal P$, by contrast, resolves the neighborhood of spatial infinity. Its physical radial interval is time-dependent, $0\le R\le R_{\rm b}(T)$. The Penrose interval expands with unit speed for $T<0$, reaches $R=\pi$ at $T=0$, and contracts with unit speed for $T>0$. At the two matching times $T=\pm\pi/2$, one has $R_{\rm b}(T)=\pi/2$, so that
\begin{equation}
  0\le R\le\pi/2
  \qquad\Longleftrightarrow\qquad
  0\le\rho\le\pi/2.
\end{equation}
The choice of the radial coordinate facilitates matching across the domains without radial coordinate interpolation. The interfaces between the three regions are the hypersurfaces
\begin{equation}
  \Sigma_- :=
  \{(\tau_-,\rho)\in\mathcal H_-:\tau_-=0\},
  \qquad
  \Sigma_{\rm P}^- :=
  \{(T,R)\in\mathcal P:T=-\pi/2\},
\end{equation}
and
\begin{equation}
  \Sigma_{\rm P}^+ :=
  \{(T,R)\in\mathcal P:T=\pi/2\},
  \qquad
  \Sigma_+ :=
  \{(\tau_+,\rho)\in\mathcal H_+:\tau_+=0\}.
\end{equation}
The identifications $\Sigma_- \simeq \Sigma_{\rm P}^-$ and $\Sigma_{\rm P}^+ \simeq \Sigma_+$ are made using the common intrinsic coordinate $R=\rho$. Consequently, the geometric matching is exact. Any further transformation of the evolved variables is determined only by the first-order reduction of the wave equation.

\begin{figure}
    \centering
    \includegraphics[width=0.3\textwidth]{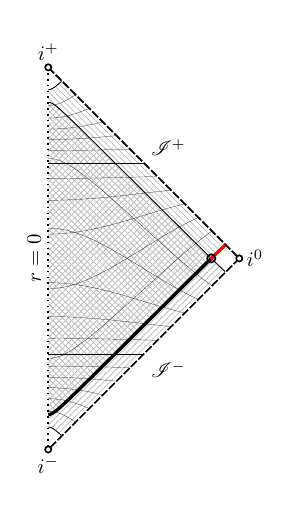}\hspace{13mm}
    \includegraphics[width=0.3\textwidth]{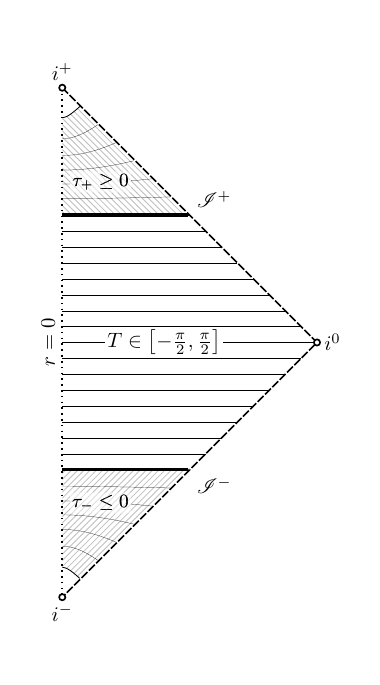}
    \caption{Comparison of strategies for bridging past null infinity $\mathscr{I}^-$ and future null infinity $\mathscr{I}^+$ in the scattering problem. Left: The method used in \cite{duarte2026numerical}, which relies exclusively on hyperboloidal foliations.  This approach yields overlapping slices and leaves an uncovered causal gap near spatial infinity $i_0$ (red line). Right: The hybrid approach proposed in this work. By combining hyperboloidal foliations near the null boundaries with Penrose coordinates around $i_0$ , the complete scattering domain is covered efficiently without artificial outer boundaries.}
    \label{fig:two_strategy}
\end{figure}

The geometric matching strategy is contrasted with the method of \cite{duarte2026numerical} in Figure~\ref{fig:two_strategy}: the purely hyperboloidal construction on the left illustrates the problem near spatial infinity and the interpolation between the hyperboloidal gauges. The right panel shows the three-region decomposition used here, with a past hyperboloidal region, a Penrose region covering $i^0$, and a future hyperboloidal region. Thus the figure provides the conformal geometric framework for the evolution $\mathcal H_- \to \mathcal P \to \mathcal H_+$, with exact matching at the interfaces $T=\pm\pi/2$.

\section{Conformal wave equation}\label{sec:conformal-semilinear}

To test the geometric framework presented in the previous section numerically, we consider a scalar wave equation on Minkowski spacetime $(\mathcal{M},\eta)$ with a general (possibly nonlinear) source term $S(x,\phi)$:
\begin{equation}
  \Box_\eta \phi = - S(x,\phi),
  \label{eq:physical-generalV}
\end{equation}
where $\Box_\eta := \eta^{ab}\nabla_a\nabla_b$ and $\phi$ is a real scalar field. We solve this equation with respect to the conformal metric $g = \Omega^2 \eta$, with a smooth conformal factor $\Omega>0$ in the physical region and $\Omega=0$ at null infinity. The conformally covariant wave (Yamabe) operator in four spacetime dimensions is
\begin{equation}
  P_g := \Box_g - \frac{1}{6}R[g],
  \label{eq:yamabe-operator}
\end{equation}
where $R[g]$ denotes the scalar curvature of $g$. We use the standard transformation law with $R[\eta]=0$:
\begin{equation}
  \Box_\eta \phi = P_\eta \phi
  = \Omega^3 P_g(\Omega^{-1}\phi)
  = -S(x,\phi).
  \label{eq:conformal-covariance}
\end{equation}
For the conformally rescaled field $\psi := \Omega^{-1}\phi$, the physical equation
\eqref{eq:physical-generalV} is equivalently written as
\begin{equation}
  P_g \psi = -\Omega^{-3}S(x,\Omega\psi).
  \label{eq:conformal-wave-source}
\end{equation}
This conformal equation is regular at null infinity for a right hand side with a smooth limit as $\Omega\to0$. We now list two examples satisfying this condition.


\paragraph{Example 1 (focusing semilinear power).}
Let
\begin{equation}\label{eq:semilinear_source}
  S_{\rm sl}(x,\phi) := |\phi|^{p-1}\phi .
\end{equation}
Then
\begin{equation}
  -\Omega^{-3}S_{\rm sl}(x,\Omega\psi)
  =
  -\Omega^{-3}|\Omega\psi|^{p-1}\Omega\psi
  =
  -\Omega^{p-3}|\psi|^{p-1}\psi .
\end{equation}
Hence
\begin{equation}
  \left(\Box_g - \frac{1}{6}R[g]\right)\psi
  =
  -\Omega^{p-3}|\psi|^{p-1}\psi .
\end{equation}
For the semilinear power, regularity at null infinity requires $p\ge3$. The marginal case is the cubic nonlinearity $p=3$, which is conformally invariant in $3+1$ dimensions. We find that this is the most challenging case for the numerical implementation due to the non-vanishing source terms at the conformal boundary (see Sec.~\ref{subsec:nonlinear_wave}).

\paragraph{Example 2 (P\"oschl--Teller potential).}
For a localized, positive, linear, spherically symmetric potential we write,
\begin{equation}
  S_U(x,\phi) := -U(r)\phi .
\end{equation}
The conformal right-hand side is
\begin{equation}
  -\Omega^{-3}S_U(r,\Omega\psi)
  =
  \Omega^{-2}U(r)\psi .
\end{equation}
Regularity at null infinity requires $U(r)$ to decay sufficiently fast as $r\to\infty$. In particular, decay with $r^{-2}$ gives a bounded conformal coefficient. The P\"oschl--Teller potential decays exponentially and gives a vanishing limit at $\mathscr I$ with
\begin{equation}
  U_{\rm PT}(r)
  :=
  V_0\,\sech^2(\alpha r)
  =
  \frac{V_0}{\cosh^2(\alpha r)} .
\end{equation}
Therefore, we get
\begin{equation}
  \left(\Box_g - \frac{1}{6}R[g]\right)\psi
  =
  \Omega^{-2}V_0\,\sech^2(\alpha r)\,\psi ,
\end{equation}
where $r$ is expressed in the relevant compactified coordinates.

\subsection{Conformal wave equation in Penrose coordinates}\label{sec:penrose-semilinear}
The conformal Penrose metric (\ref{eq:penrose_metric}) describes the Einstein static universe with scalar curvature $R[g]=6$. We have $\sqrt{|g|} = \sin^2R\,\sqrt{|\sigma|}$ and we write \(\Delta_\sigma\) to denote the Laplace--Beltrami operator of the
unit round metric $\sigma$ on $\mathbb S^2$, with the convention $\Delta_\sigma Y_{\ell m}=-\ell(\ell+1)Y_{\ell m}$. The explicit Penrose conformal equation becomes
\begin{equation}\label{eq:conformal-semilinear-penrose-explicit}
\hspace{-1cm}
-\partial_T^2\psi + \partial_R^2\psi + 2\cot R\,\partial_R\psi + \frac{1}{\sin^2R}\Delta_\sigma\psi - \psi
=
-\left(\cos T+\cos R\right)^{p-3}\,|\psi|^{p-1} \psi.
\end{equation}
Note that the left hand side is independent of $T$. The physical domain, however, is time-dependent because the conformal boundary is given by $\Omega_P=0$, i.e.
\[
  \cos T + \cos R = 0
  \qquad \Longleftrightarrow \qquad
  R = R_{\rm b}(T) = \pi-|T|.
\]
We perform the Penrose evolution on the fixed numerical interval $R\in[0,\pi]$, but the physical Minkowski region occupies the moving subinterval $0\le R\le R_{\rm b}(T)$. In the past half of the Penrose evolution ($T<0$), this boundary moves outward with unit speed, and in the future half ($T>0$) it moves inward with unit speed. The Penrose conformal factor is
\begin{equation}\label{eq:conformal_penrose}
  \Omega_P = \cos T + \cos R,
\end{equation}
and the physical Minkowski radius is
\begin{equation}
  r(T,R) = \frac{\sin R}{\Omega_P} = \frac{\sin R}{\cos T+\cos R}.
\end{equation}
For the semilinear source term:
\begin{equation}\label{eq:rhs-semilinear-penrose}
  S_{\rm sl}(r,\phi) = |\phi|^{p-1}\phi
  \qquad\Longrightarrow\qquad
  \rm{RHS} = -\Omega_P^{p-3}\,|\psi|^{p-1} \psi.
\end{equation}
For the P\"oschl--Teller test we take a linear potential of the form
\begin{equation}\label{eq:pt-potential}
  S_{\rm PT}(r,\phi) = -U_{\rm PT}(r)\,\phi,
  \qquad
  U_{\rm PT}(r) = \frac{V_0}{\cosh^2(\alpha r)}.
\end{equation}
Then we have
\begin{equation}\label{eq:rhs-pt-penrose}
  \rm{RHS}
  =
  \Omega_P^{-2}\,U_{\rm PT}(r)\,\psi
  =
  (\cos T+\cos R)^{-2}\,
  \frac{V_0}{\cosh^2 \left(\alpha\,\frac{\sin R}{\cos T+\cos R}\right)}\,\psi.
\end{equation}

\begin{figure}[ht]
    \centering
    \includegraphics[width=0.65\linewidth]{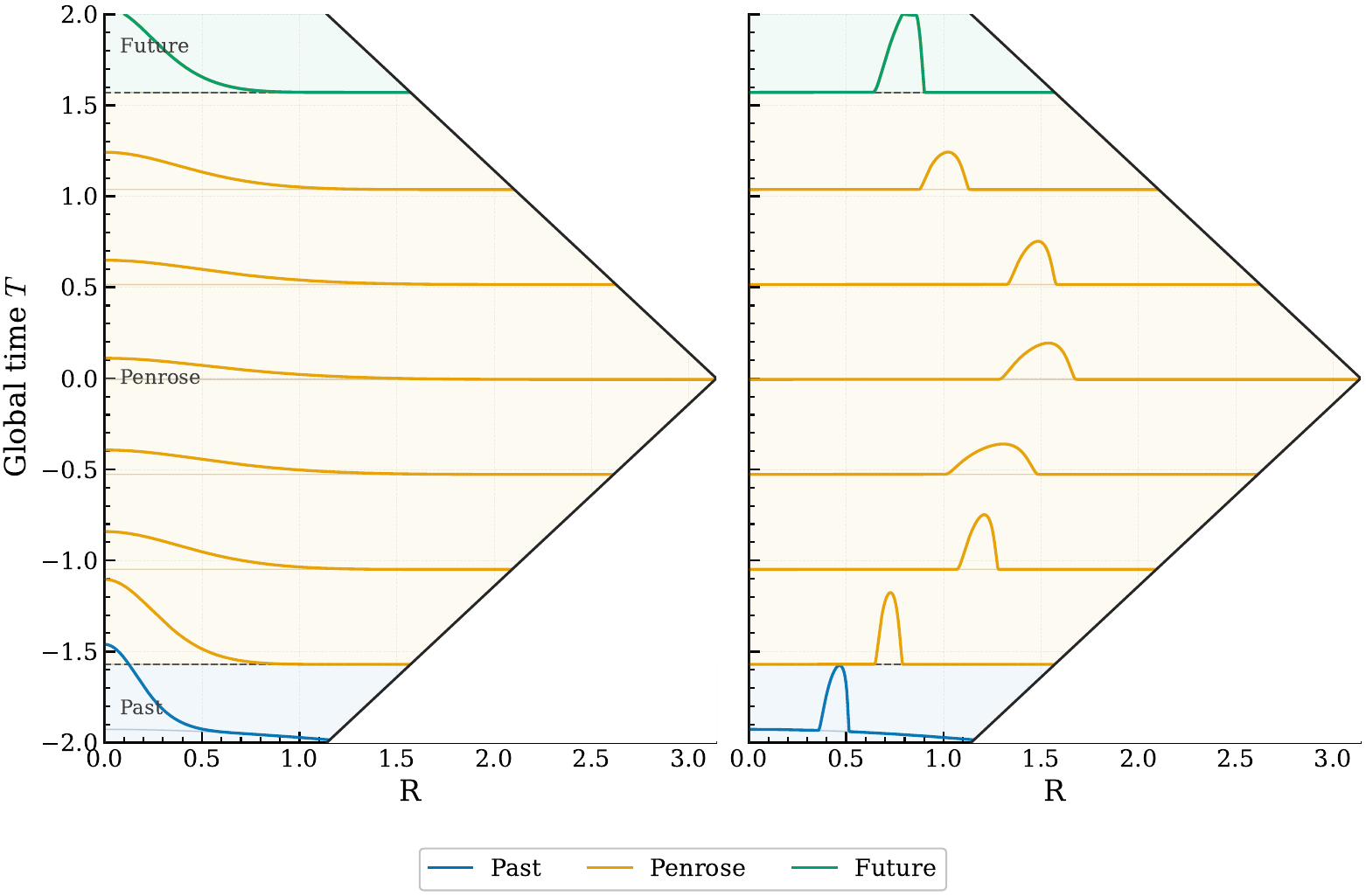}
    \caption{Snapshots of the potentials used in this study in different coordinate systems. Left: P\"oschl--Teller. Right: shaking barrier.}
    \label{fig:potentials}
\end{figure}

The time-dependent Penrose conformal factor in \eref{eq:conformal_penrose} induces a time dependence in otherwise static potentials. In Figure~\ref{fig:potentials}, we plot snapshots of the P\"oschl--Teller potential and of the localized time-dependent barrier adapted from \cite{duarte2026numerical} and used in the numerical experiments below.

\subsection{Conformal wave equation in stationary hyperboloidal coordinates}\label{subsec:RHSandpotentials}
We now specialize to the hyperboloidal compactification as defined in (\ref{eq:past_hyperboloidal}) and (\ref{eq:future_hyperboloidal}) and the conformal metric (\ref{eq:hyperboloidal_metric}). The scalar curvature is
\begin{equation}\label{eq:hyperboloidal-curvature}
  R[g] = 18\cos^2\rho,
\end{equation}
Denoting by $\Delta_\sigma$ the Laplacian on $\mathbb S^2$, the hyperboloidal conformal equation reads
\begin{equation}
\label{eq:conformal-semilinear-hyperboloidal}
\hspace{-1cm}
\begin{array}{@{}l@{}}
  -\partial_{\tau_\pm}^2\psi
  \mp 2\sin\rho\,\partial_{\tau_\pm}\partial_\rho\psi
  +\cos^2\rho\,\partial_\rho^2\psi
  \mp 3\cos\rho\,\partial_{\tau_\pm}\psi
  +2\cot\rho\,\cos(2\rho)\,\partial_\rho\psi
  \\[0.4em]
  \qquad
  +\frac{1}{\sin^2\rho}\,\Delta_\sigma\psi
  -3\cos^2\rho\,\psi
  =
  -(\cos\rho)^{p-3}\,|\psi|^{p-1} \psi.
\end{array}
\end{equation}
The upper sign corresponds to the future hyperboloidal evolution with $\tau_+$, and the lower sign to the past hyperboloidal evolution with $\tau_-$. In our experiments we consider the spherically symmetric solutions and set $\Delta_\sigma\psi=0$.

\paragraph{Right-hand sides (semilinear and P\"oschl--Teller).}
In these coordinates the conformal factor is
\begin{equation}
  \Omega_H = \cos\rho,
\end{equation}
and (using the compactification in (\ref{eq:past_hyperboloidal}) and (\ref{eq:future_hyperboloidal})) the physical radius is
\begin{equation}
  r = \tan\rho.
\end{equation}
For the semilinear source term:
\begin{equation}\label{eq:rhs-semilinear-hyperboloidal}
  S_{\rm sl}(r,\phi) = |\phi|^{p-1} \phi
  \qquad\Longrightarrow\qquad
  \mathrm{RHS} = -(\cos\rho)^{p-3}\,|\psi|^{p-1} \psi.
\end{equation}
For the P\"oschl--Teller test with \eref{eq:pt-potential}, 
\begin{equation}\label{eq:rhs-pt-hyperboloidal}
  \mathrm{RHS}
  =
  \Omega_H^{-2}\,U_{\rm PT}(r)\,\psi
  =
  (\cos\rho)^{-2}\,
  \frac{V_0}{\cosh^2 \left(\alpha\,\tan\rho\right)}\,\psi.
\end{equation}
\section{Numerical Implementation}\label{sec:numerics}

\subsection{Equations and data}\label{sec:eqs_data}

For the numerical implementation, we reduce the systems to first-order form. The first-order reduction variables introduced below are denoted by $\Psi$ and $\Pi$. Assuming spherical symmetry, the conformal metric takes the general form
\begin{equation}
\bar{g}_{\mu\nu} \, \rm{d}x^\mu \rm{d}x^\nu = 
(-\alpha^2 + \beta^2 \gamma^2) \, \rm{d}\tau^2 + 2\beta \gamma^2 \, \rm{d}\tau \, \rm{d}s
+ \gamma^2 \, \rm{d}s^2 + \lambda^2 \, \rm{d}\sigma^2,
\end{equation}
where $\tau$ is the patch time coordinate, $s$ is the corresponding compactified radial coordinate ($s=\rho$ in the hyperboloidal regions and $s=R$ in the Penrose region), and $\lambda$ is the angular scale factor. In our coordinates, $\lambda=\sin\rho$ in the hyperboloidal patches and $\lambda=\sin R$ in the Penrose patch. Following \cite{bizon2009universality}, we define the auxiliary variables
\begin{equation}
\Psi := \partial_s \psi, \qquad \Pi := \frac{\gamma}{\alpha} \left( \partial_\tau \psi - \beta \partial_s \psi \right).
\label{eq:first-order-variables}
\end{equation}
In each patch we write the conformal wave equation in the form
\begin{equation}
  \left(\Box_{\bar{g}}-\frac{1}{6}\bar{R}\right)\psi = -S(\tau,x,\psi),
\end{equation}
where all model dependence is contained in the conformal source term $S$. These definitions reduce the equation to the first-order symmetric hyperbolic system
\begin{equation}
\label{eq:first-order-system}
\begin{array}{rcl}
\partial_\tau \psi &=& \displaystyle \frac{\alpha}{\gamma} \Pi + \beta \Psi,\\[0.5em]
\partial_\tau \Psi &=& \displaystyle \partial_s\left(\frac{\alpha}{\gamma} \Pi + \beta \Psi\right),\\[0.5em]
\partial_\tau \Pi &=&
\displaystyle \frac{1}{\lambda^2}\partial_s\left[
\lambda^2\left(\frac{\alpha}{\gamma}\Psi+\beta\Pi\right)
\right]
-\alpha\gamma\left(\frac{1}{6}\bar R\,\psi - S(\tau,x,\psi)\right).
\end{array}
\end{equation}

For the numerical experiments in this paper, the same evolution system is used in every region and for every model; only the choice of $S$ changes. For the free conformal wave equation, we take
\begin{equation}
  S(\tau,x,\psi)=0.
\end{equation}
For the semilinear model with $S_{\rm sl}(\phi)=|\phi|^{p-1} \phi$, the source becomes
\begin{equation}
  S_{\rm sl}(\tau,x,\psi) = \Omega^{p-3} |\psi|^{p-1} \psi.
\end{equation}
For the linear localized-potential models of Section~\ref{sec:results}, defined physically by $\Box_\eta\phi=V(t,r)\phi$, the corresponding source in the first-order convention above is
\begin{equation}
  S_{\rm pot}(\tau,x,\psi) = -\Omega^{-2} V\left(t(\tau,x),r(\tau,x)\right)\psi.
\end{equation}
Here $\Omega$ denotes the patch-dependent conformal factor, namely $\Omega_H$ in the hyperboloidal patches and $\Omega_P$ in the Penrose patch. This includes the P\"oschl--Teller profile as the special case $V(t,r)=U_{\rm PT}(r)$.

The incoming data are defined at $\mathscr{I}^-$. In the numerical studies presented below, we use the Gaussian radiation profile
\begin{equation}\label{eq:initial_gaussian}
    G(\tau_-) = A e^{- {(\tau_- - \tau_{0})^2}/{2\sigma^2} },
    \qquad
    G'(\tau_-) = - \frac{(\tau_- - \tau_0)}{\sigma^2}  G(\tau_-).
\end{equation}
Unless stated otherwise, the simulations use $A=1$, $\sigma=0.05$,  and $\tau_0=-3$. The initial past-hyperboloidal slice is placed at $\tau_{-,0}=-6$. With these parameters the Gaussian is negligible both on the initial slice and at the matching surface $\tau_-=0$. Although the profile is Gaussian rather than exactly compactly supported, the parameters are chosen so that the data are below machine precision at the transition.

The bulk initial data on the first past-hyperboloidal slice are taken to vanish:
\begin{equation}
  \psi(\tau_{-,0},\rho)=0,\qquad
  \Psi(\tau_{-,0},\rho)=0,\qquad
  \Pi(\tau_{-,0},\rho)=0,
  \qquad 0\le \rho < \frac{\pi}{2}.
\end{equation}
The subsequent evolution is therefore driven entirely by the prescribed characteristic boundary data at $\mathscr{I}^-$ derived from (\ref{eq:initial_gaussian}); the precise characteristic prescription is given in Section~\ref{subsec:boundary-conditions}. The incoming signal is attached to null infinity itself rather than to an arbitrarily chosen initial slice, thereby eliminating slicing dependence from the prescribed input. What remains is only the choice of parametrization along $\mathscr{I}^-$.

This system is symmetric hyperbolic and suitable for numerical evolution using high-order finite differences and explicit time integration. Table~\ref{tab:coefficients} lists the values of the lapse $\alpha$, shift $\beta$, spatial metric function $\gamma$, conformal factor ($\Omega_H$ in the hyperboloidal patches and $\Omega_P$ in the Penrose patch), angular scale $\lambda$, and scalar curvature $\bar{R}$.

\begin{table}[ht]
    \centering
    \caption{Metric coefficients in different coordinate patches.}
    \label{tab:coefficients}
    \begin{tabular}{lcccccc}
    \hline
    Region & Conformal factor & $\alpha$ & $\gamma$ & $\beta$ & $\lambda$ & $\bar{R}$\\
    \hline
    Past hyperboloidal & $\Omega_H=\cos \rho$ & 1 & 1 & $\sin \rho$ & $\sin \rho$ & $18\cos^2 \rho$\\
    Future hyperboloidal & $\Omega_H=\cos \rho$ & 1 & $1$ & $-\sin \rho$ & $\sin \rho$ & $18\cos^2 \rho$\\
    Penrose & $\Omega_P=\cos R + \cos T$ & 1 & 1 & 0 & $\sin R$ &  6\\

    \hline
    \end{tabular}
\end{table}

The numerical solver proceed as follows:
\begin{enumerate}
\item Prescribe vanishing initial data on the past hyperboloidal slice $\tau_-=\tau_{-,0}=-6$ in $\mathcal{H}_-$ with regularity condition imposed at $\rho=0$.

\item Evolve the past hyperboloidal system on $\mathcal{H}_-$ forward in $\tau_-$ with incoming boundary data at $\scri^-$ up to the matching hypersurface $\Sigma_-$.

\item Transfer the data from $\Sigma_-$ to $\Sigma_{\rm P}^-$ using the pointwise identification $R=\rho$ and the transformation law for $U$ given in Sec.~\ref{sec:interface}.

\item Evolve the Penrose system on $\mathcal{P}$ for $-\pi/2 \le T \le \pi/2$.

\item Transfer the data from $\Sigma_{\rm P}^+$ to $\Sigma_+$ using the pointwise identification $\rho=R$ and the transformation law for $U$ given in Sec.~\ref{sec:interface}.

\item Evolve the future hyperboloidal system on $\mathcal{H}_+$ forward in $\tau_+$.

\item Read off the outgoing radiation at $\rho=\pi/2=\mathscr{I}^+$ and compute the scattering observables.
\end{enumerate}

From a numerical point of view, the main advantage of this decomposition is that each region is adapted to the geometry it resolves best: the hyperboloidal patches provide fixed outer boundaries at null infinity suitable for long-time simulations, whereas the Penrose patch enables the transition across spatial infinity. The interfaces at $T=\pm \pi/2$ are particularly simple because the physical Penrose domain there coincides exactly with the hyperboloidal radial interval. As a consequence, the passage between $\mathcal{H}_-$, $\mathcal{P}$, and $\mathcal{H}_+$ is implemented by pointwise transfer of the state vector, with the only nontrivial component being the conformally induced corrections \eqref{eq:past-interface-transfer} and \eqref{eq:future-interface-transfer} for $\Pi$. This yields a clean and geometrically natural realization of the time-domain scattering map $ \mathscr{I}^- \to \mathscr{I}^+$. Next, we describe the transition algorithm across the patch interfaces in detail.

\subsection{Patch interfaces}\label{sec:interface}
The patching step uses the geometric matching described in Section~\ref{sec:penrose-hyperboloidal-matching}. In particular, \eref{eq:rho-equals-R-on-matching} shows that the hypersurfaces $\tau_-=0$ and $T=-\pi/2$, and likewise $T=\pi/2$ and $\tau_+=0$, are the same physical hypersurfaces with common intrinsic coordinate $R=\rho$. The active radial intervals therefore coincide at both interfaces, so the numerical transfer is pointwise and requires no radial interpolation.

Since $\Omega_{\rm P}=\Omega_H=\cos\rho$ on the matching surfaces, the conformal field and its tangential derivative are copied directly:
\begin{equation}
\begin{array}{lll}
  \psi_{\rm P}=\psi_-, & \Psi_{\rm P}=\Psi_-,
  & \hbox{on } \Sigma_-\simeq\Sigma_{\rm P}^-,\\[2mm]
  \psi_+=\psi_{\rm P}, & \Psi_+=\Psi_{\rm P},
  & \hbox{on } \Sigma_{\rm P}^+\simeq\Sigma_+ .
\end{array}
\end{equation}
Only $\Pi$ receives a correction, because its definition \eref{eq:first-order-variables} involves the evolution direction and hence the normal derivative of the conformal factor. Using the coefficients in Table~\ref{tab:coefficients}, the common derivative at the past interface is $\partial_{\tau_-}-\sin\rho\,\partial_\rho=\partial_T$, while at the future interface it is $\partial_{\tau_+}+\sin\rho\,\partial_\rho=\partial_T$. These identities follow directly from the coordinate maps in \eref{eq:rho-equals-R-on-matching}.

For two conformal factors that agree on an interface, $\psi_A=\Omega_A^{-1}\phi$ and $\psi_B=\Omega_B^{-1}\phi$ satisfy
\begin{equation}
  D\psi_A
  =
  D\psi_B
  -
  \psi\,\frac{D\Omega_A-D\Omega_B}{\Omega},
  \qquad
  \Omega_A=\Omega_B=\Omega .
  \label{eq:interface-conformal-derivative}
\end{equation}
At the past interface $(D\Omega_{\rm P}-D\Omega_H)/\Omega_H=\cos\rho$, whereas at the future interface this factor is $-\cos\rho$. The resulting transfer law from the past hyperboloidal patch to the Penrose patch is
\begin{equation}
  \psi_{\rm P}=\psi_-,
  \qquad
  \Psi_{\rm P}=\Psi_-,
  \qquad
  \Pi_{\rm P}=\Pi_- - \cos\rho\,\psi_-,
  \qquad R=\rho .
  \label{eq:past-interface-transfer}
\end{equation}
The transfer from the Penrose patch to the future hyperboloidal patch is
\begin{equation}
  \psi_+=\psi_{\rm P},
  \qquad
  \Psi_+=\Psi_{\rm P},
  \qquad
  \Pi_+=\Pi_{\rm P} - \cos R\,\psi_{\rm P},
  \qquad \rho=R .
  \label{eq:future-interface-transfer}
\end{equation}

Thus the implemented patch transition consists only of copying $\psi$ and $\Psi$ at common grid points, applying \eref{eq:past-interface-transfer} or \eref{eq:future-interface-transfer} to $\Pi$, and continuing the time update with the coefficients of the receiving patch. The treatment of the fixed Penrose grid, its active physical interval, and the endpoint regularity at $R=\pi$ is described in Sections~\ref{subsec:spatial-discretization} and \ref{subsec:regularity-origin-i0}.

\subsection{Spatial discretization and time integration}
\label{subsec:spatial-discretization}

We solve the symmetric hyperbolic reduction of the conformal wave equation \eqref{eq:first-order-system} with the coefficients given in Table~\ref{tab:coefficients} using fourth-order finite differences in space and fourth-order Runge--Kutta integration in time. The time step is determined by a Courant-Friedrichs-Lewy (CFL) condition based on the maximum characteristic speed, with the CFL factor set to 0.5. The spatial derivative operator is centered in the interior and one-sided at the boundaries. The regularity conditions at the origin and at $R=\pi$, described in Section~\ref{subsec:regularity-origin-i0}, are constructed to be consistent with this fourth-order spatial discretization.

To suppress spurious high-frequency numerical noise arising from grid boundaries and nonlinear interactions, we add a Kreiss--Oliger (KO) dissipation term to the right-hand side of all evolved variables $u=(\psi,\Psi,\Pi)$. The discrete operator added to each evolution equation is
\[
D_{{KO}}[u_A] = \sigma_{{KO} }\frac{h^5}{64} (D_+ D_-)^3 u_A,
\qquad u_A\in\{\psi,\Psi,\Pi\},
\]
where $h$ is the spatial grid resolution, $D_+$ and $D_-$ are the standard forward and backward finite difference operators, and the dissipation parameter is set to $\sigma_{{KO} } = 0.1$.

The past and future hyperboloidal regions are discretized on uniform grids in $\rho\in[0,\pi/2]$, so their outer boundaries remain fixed at null infinity. The Penrose region is discretized on a uniform grid in $R\in[0,\pi]$. Its numerical outer edge is fixed, but the physical Minkowski domain at time $T$ is only the active subinterval $0\le R\le R_{\rm b}(T)=\pi-|T|$. Therefore, the physical Penrose boundary moves across the fixed grid, advancing outward with unit speed for $T<0$ and retreating inward with unit speed for $T>0$. Grid points outside the physical domain with $R>R_{\rm b}(T)$ are excluded from diagnostics. For nonlinear evolutions the nonlinear source is evaluated only where $\Omega_P>0$, preventing the inactive part of the fixed grid from contributing to the physical source term.

\subsection{Regularity at the origin and spatial infinity}
\label{subsec:regularity-origin-i0}
The angular scale factor $\lambda$ vanishes at the center of spherical symmetry and also at the right endpoint of the Penrose grid. In the hyperboloidal patches $\lambda=\sin\rho$ vanishes at $\rho=0$, while in the Penrose patch $\lambda=\sin R$ vanishes at both $R=0$ and $R=\pi$. The point $R=\pi$ is spatial infinity $i^0$ at $T=0$. Numerically, $R=\pi$ is therefore treated as an endpoint of the compactified radial coordinate, not as an outer boundary at which physical data are prescribed.

The potentially singular term in the $\Pi$ equation is evaluated in a form that makes this endpoint regularity explicit. Defining
\begin{equation}
  Z := \frac{\alpha}{\gamma}\Psi+\beta\Pi,
  \qquad
  F := \frac{Z}{\lambda},
\end{equation}
the radial principal part is written as
\begin{equation}
  \frac{1}{\lambda^2}\partial_x\left(\lambda^2 Z\right)
  =
  3\,\partial_x Z - 2\lambda\,\partial_x F .
  \label{eq:regularized-radial-principal-part}
\end{equation}
This identity is used directly in the code. The quotient $F=Z/\lambda$ is formed only away from points where $\lambda=0$. At the degenerate endpoints, the implementation replaces the endpoint value of $F$ by its regular limiting value obtained from l'Hôpital's rule. Explicitly,
\begin{equation}
  F(0)
  =
  \lim_{x\to0}\frac{Z}{\lambda}
  =
  \left.\partial_x Z\right|_{x=0},
  \qquad
  F(\pi)
  =
  \lim_{R\to\pi}\frac{Z}{\sin R}
  =
  -\left.\partial_R Z\right|_{R=\pi}.
  \label{eq:lhopital-endpoint-values}
\end{equation}
The first formula applies at the origin, where $x=\rho$ in the hyperboloidal patches and $x=R$ in the Penrose patch. The second formula applies only at the right endpoint of the Penrose grid and carries the minus sign because $\partial_R\sin R=-1$ at $R=\pi$.

At the origin, smooth spherical symmetry implies that $\psi$ and $\Pi$ are even functions of the local radial coordinate, while $\Psi=\partial_x\psi$ and $Z$ are odd. The code enforces this regularity in differential form by setting the right-hand side of the $\Psi$ equation to zero at the origin. For the $\Pi$ equation, the endpoint value is set by the limit of \eref{eq:regularized-radial-principal-part}; since $\lambda=0$ at the endpoint, the term proportional to $\lambda\,\partial_xF$ drops out there. The first few neighboring points are then recomputed using one-sided finite-differences in which the missing endpoint value of $F$ is replaced by the corresponding l'Hôpital limit. The Penrose endpoint $R=\pi$ is handled by the same mechanism.

This regularity treatment is separate from the moving null boundary $R=R_{\rm b}(T)$. The finite-difference evolution in the Penrose phase is performed on the fixed interval $0\le R\le\pi$, while physical diagnostics are restricted to $0\le R\le R_{\rm b}(T)$. 


\subsection{Diagnostics}\label{sec:diagnostics}
To verify the accuracy of the numerical scheme, we define the convergence factor $Q$ by comparing solutions at different resolutions. Given numerical solutions $\psi_h$, $\psi_{h/2}$, and $\psi_{h/4}$ computed with grid spacings $h$, $h/2$, and $h/4$ respectively, the convergence factor is defined as
\begin{equation}\label{eq:convergence_factor}
Q_h = \log_2\left( \frac{||\psi_h - \psi_{h/2}||}{||\psi_{h/2} - \psi_{h/4}||} \right),
\end{equation}
where $||\cdot||$ denotes the discrete $L^2$ norm over the computational grid.

To analyze the asymptotic behavior and late-time radiation tails of the fields, we define the local power index
\begin{equation}
p_\rho(\tau_+) = \frac{d\ln|\psi(\tau_+,\rho)|}{d\ln \tau_+},
\end{equation}
If $\psi\sim t^{-q} \sim \tau_+^{-q}$, then $p_\rho(\tau_+)\to -q$. For semilinear wave equations, these late-time tail decay rates are known. Specifically, for the cubic nonlinearity, the field is expected to decay as $t^{-1}$ at future null infinity $\mathscr{I}^+$ and as $t^{-2}$ at the origin, corresponding to asymptotic power indices of $-1$ and $-2$, respectively~\cite{bizon2009universality,Donninger_2014}. Similarly, the expected indices for the quintic equation are $-3$ at $\mathscr{I}^+$, and $-4$ at the origin while the septic equation yields theoretical rates of $-5$ and $-6$~\cite{Rinne_2025}. We compute this local power index to directly compare our numerical late-time tails against these expected analytical decay rates in Section~\ref{subsec:nonlinear_wave}.

\subsection{Boundary data}
\label{subsec:boundary-conditions}
For the first-order system, boundary data are imposed on characteristic fields. The principal part of the $(\Psi,\Pi)$ subsystem has characteristics
\begin{equation}
  C_{\leftarrow} := \Pi+\Psi,\qquad
  C_{\rightarrow} := \Pi-\Psi,
\end{equation}
The characteristic speeds are
\begin{equation}
  v_{\leftarrow}
  =
  -\left(\beta+\frac{\alpha}{\gamma}\right),
  \qquad
  v_{\rightarrow}
  =
  -\left(\beta-\frac{\alpha}{\gamma}\right).
\end{equation}
These fields determine which boundary data may be prescribed.

On the past hyperboloidal patch, the outer boundary $\rho=\pi/2$ coincides with $\mathscr I^-$. Since $\beta=1$ and $\alpha/\gamma=1$ there, $C_{\leftarrow}$ has speed $-2$ and enters the domain, while $C_{\rightarrow}$ is tangent to the boundary. The incoming Gaussian profile $G$ in \eref{eq:initial_gaussian} is therefore imposed through
\begin{equation}
  C_{\leftarrow}(\tau_-,\pi/2)=G'(\tau_-).
\end{equation}
In the implementation this is enforced at every Runge--Kutta substep by setting
\begin{equation}
  \Pi(\tau_-,\pi/2)=\Psi(\tau_-,\pi/2)=\frac{1}{2}G'(\tau_-).
\end{equation}
Consequently $C_{\leftarrow}=G'$, $C_{\rightarrow}=0$, and the boundary evolution equation gives $\partial_{\tau_-}\psi=\Pi+\Psi=G'$. The boundary value $\psi(\tau_-,\pi/2)=G(\tau_-)$ is obtained by time integration from compatible initial data; it is not imposed as an additional independent Dirichlet condition. The initial corner compatibility condition is $\psi(\tau_{-,0},\pi/2)=G(\tau_{-,0})$. For the choices $\tau_{-,0}=-6$, $\tau_0=-3$, and $\sigma=0.05$, this value is below machine precision.

On the future hyperboloidal patch, the same outer grid point represents $\mathscr I^+$, but now $\beta=-1$. Therefore, $C_{\rightarrow}$ has positive speed and leaves the domain, while $C_{\leftarrow}$ is tangent to $\mathscr I^+$. No incoming characteristic field is present and no boundary data are imposed; the outgoing radiation is simply read off at the boundary.

The Penrose patch uses a fixed grid $0\le R\le\pi$, while the physical conformal boundary is the moving curve $R=R_{\rm b}(T)=\pi-|T|$. Here $\beta=0$, so $C_{\leftarrow}$ and $C_{\rightarrow}$ have speeds $-1$ and $+1$, respectively. Since the simulation uses a fixed spatial grid $R\in[0,\pi]$, we do not apply boundary conditions beyond the regularity conditions described in Sec.~\ref{subsec:regularity-origin-i0}.


\section{Results}\label{sec:results}
To validate the global numerical framework, we first consider the free propagation of scalar waves. We establish the accuracy of our code by evolving the free conformal wave equation and comparing it against an exact global analytical solution detailed in \ref{sec:free-test-solution}. We then investigate wave scattering in the presence of the P\"oschl--Teller potential and the localized time-dependent potentials studied in~\cite{duarte2026numerical}. Finally, we extend the framework to evaluate nonlinear scattering, analyzing the late-time radiation tails of semilinear wave equations.

These numerical experiments demonstrate the practical effectiveness of our spatial compactification framework. The compactified multi-domain construction avoids artificial timelike outer boundaries and allows the scattering data to be prescribed and measured directly at null infinity.

\subsection{The free wave equation} \label{sec:free_wave}

The simplest test of the global evolution scheme is the propagation of a free, spherically symmetric wave governed by $\Box_\eta \phi = 0$. We drive the evolution using the Gaussian pulse formulation prescribed at past null infinity, $\mathscr{I}^-$.

To accurately resolve this evolution across the central Penrose patch, the numerical scheme must cleanly isolate the physical Minkowski domain from the inactive part of the fixed Penrose grid. As detailed in Section~\ref{sec:numerics}, the physical domain occupies a time-dependent subinterval of the fixed numerical grid. Fields outside this active interval are excluded from diagnostics and are actively damped to avoid contamination into the physical domain.

With this boundary treatment, the pulse propagates smoothly across the domain in the absence of nonlinearities, preserving its shape and amplitude. Capturing this exact behavior serves as a nontrivial validation of the hybrid-coordinate evolution. It demonstrates that the numerical scheme successfully handles spatial compactification, grid patching, and the transition through spatial infinity without introducing visible numerical distortions or artificial reflections.

\begin{figure}[ht]
    \centering
    \includegraphics[width=0.4\linewidth]{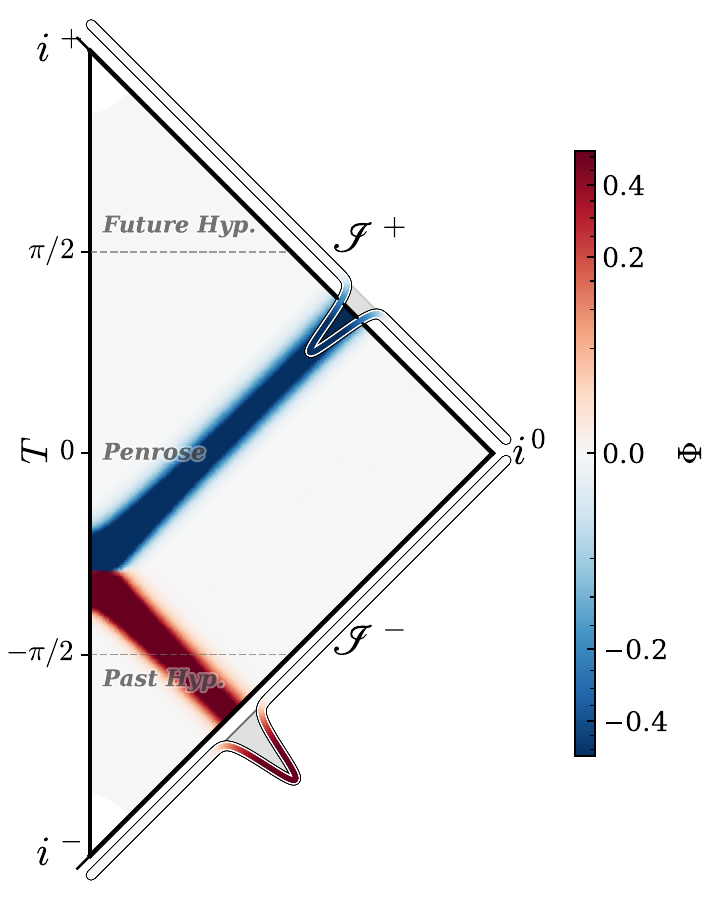}
    \includegraphics[width=0.4\linewidth]{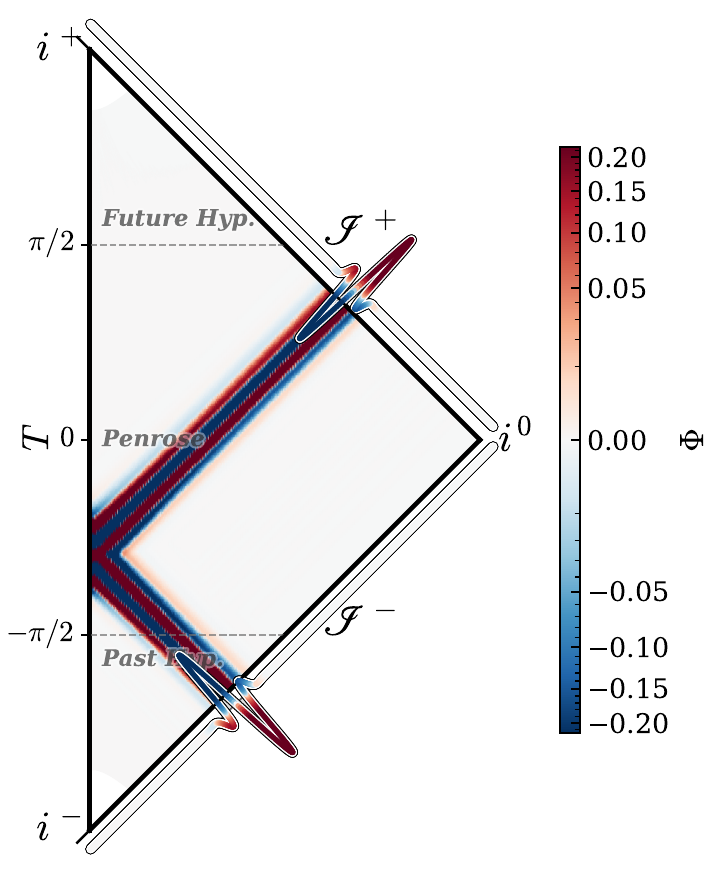}
    
    \caption{Global evolution of a free scalar wave superimposed on Penrose diagrams for two distinct incident profiles: a standard Gaussian pulse (left) and a modulated Gaussian packet $G(s) = A \sin(\omega(s - \tau_0)) \exp(-(s - \tau_0)^2/(2\sigma^2))$ (right). The plotted asymptotic profiles are the oriented radiation fields $\mathcal R_-$ on $\mathscr I^-$ and $\mathcal R_+$ on $\mathscr I^+$, defined in \eqref{eq:radiation-field-normalization}. With this normalization the free scattering map is the identity.}
    \label{fig:linear_pulse}
\end{figure}

Figure~\ref{fig:linear_pulse} illustrates the global evolution of the wave superimposed on Penrose diagrams for two distinct initial data profiles: a standard Gaussian pulse and a frequency-modulated Gaussian packet. In both cases, the initial data prescribed at past null infinity, $\mathscr{I}^-$, and the resulting outgoing radiation extracted at future null infinity, $\mathscr{I}^+$, are plotted alongside the conformal boundaries. 
At the level of the rescaled (conformal) field \(\psi \sim r\phi\), regular propagation through the origin reverses the sign of the profile. Incoming data \(G(v)\) at \(\mathscr I^-\) gives the outgoing data \(-G(u)\) at \(\mathscr I^+\). In this paper we use the oriented radiation fields
\begin{equation}\label{eq:radiation-field-normalization}
  \mathcal R_-(v)=\lim_{\mathscr I^-} w,
  \qquad
  \mathcal R_+(u)=-\lim_{\mathscr I^+} w .
\end{equation}
With this convention the free spherically symmetric scattering map is the identity, $\mathcal R_+(s)=\mathcal R_-(s)$ (compare Fig.~\ref{fig:linear_pulse}).

To verify the accuracy of our numerical scheme, we evaluate the pointwise convergence and the global order of convergence. Unlike the self-convergence test discussed in Section~\ref{sec:diagnostics}, we compute the error here by comparing the numerical results directly against the exact analytical solution derived in \ref{sec:free-test-solution}. We define the exact convergence factor as
\[
Q_{h,\mathrm{exact}} = \log_2\left( \frac{||\psi_h - \psi_{\mathrm{exact}}||}{||\psi_{h/2} - \psi_{\mathrm{exact}}||} \right).
\]

\begin{figure}[ht]
    \centering
    \includegraphics[width=0.48\textwidth]{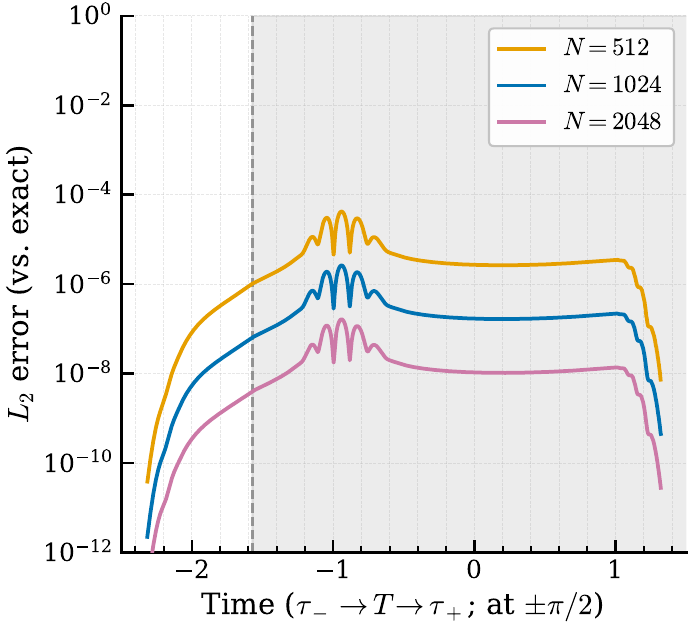}
    \includegraphics[width=0.48\textwidth]{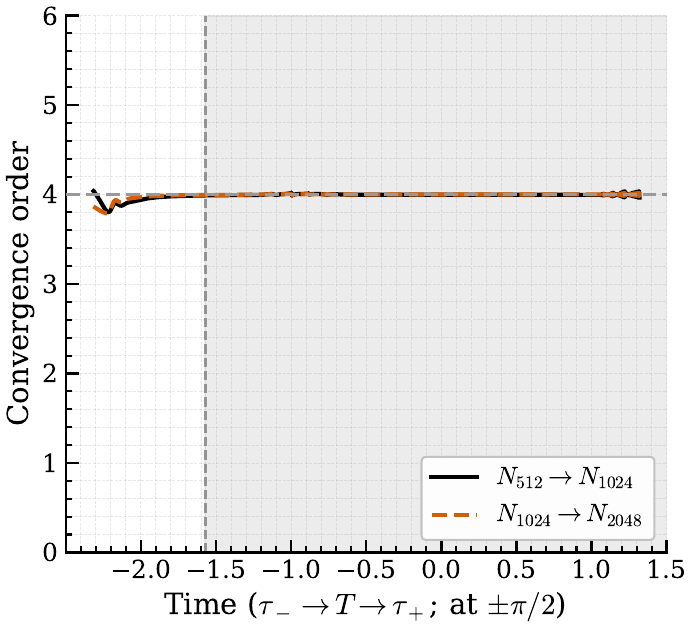}
    \caption{Convergence analysis against the exact analytical solution. \textit{Left Panel:} Time evolution of the $L_2$ error between the numerical and exact solutions of the free conformal wave equation. \textit{Right Panel:} The estimated convergence order, showing the expected fourth-order behavior across the coordinate patches.}
    \label{fig:convergence_exact}
\end{figure}

Figure~\ref{fig:convergence_exact} displays the discrete $L_2$ error for different grid resolutions across the numerical domain, together with the computed convergence factor $Q_{h,\mathrm{exact}}$. The results show fourth-order convergence relative to the exact analytical solution across the coordinate patches.

For the free wave equation, Gaussian data prescribed at $\mathscr{I}^-$ propagate strictly along null characteristics. For numerical purposes, once the Gaussian amplitude falls below machine precision, the signal is effectively compactly supported and exits $\mathscr{I}^+$ within the central Penrose patch to numerical accuracy. While this renders the future hyperboloidal region unnecessary for baseline evolution, localized potentials and nonlinearities fundamentally alter the late-time dynamics. Internal scattering generates persistent radiation tails, ensuring the outgoing field is no longer numerically compactly supported. The future hyperboloidal patch is therefore essential to resolve this radiation reaching $\mathscr{I}^+$ at arbitrarily late retarded times. We examine these scattering phenomena in the following sections.

\subsection{Wave equation with potentials}

We consider the propagation of scalar waves in the presence of localized linear background potentials. For these tests we write the physical wave equation as
\begin{equation} \label{eq:potential_wave_eqn}
\Box_{\eta}\phi = V(t,r)\phi .
\end{equation}
This sign convention gives the standard reduced radial equation $w_{tt}-w_{rr}+V(r)w=0$ for $w=r\phi$, which is the convention used below for the positive P\"oschl--Teller barrier and its quasinormal-mode spectrum. In terms of the source convention used in Section~\ref{sec:conformal-semilinear}, this corresponds to $S_{\rm pot}(t,r,\phi)=-V(t,r)\phi$. Equivalently, if the conformal equation is written as $P_{\bar g}\psi=-\mathcal S$, then the potential contribution is $\mathcal S_{\rm pot}=-\Omega^{-2}V(t,r)\psi$.

To evaluate the framework's capability to handle different scattering problems, we consider two potential families: the classic P\"oschl--Teller potential and a generalized localized barrier potential based on models recently studied by Duarte-Baptista \textit{et al.}~\cite{duarte2026numerical}. For the P\"oschl--Teller model,
\begin{equation}\label{eq:waveq_pt}
  V(t,r)=V_{\mathrm{PT}}(r),
  \qquad
  V_{\mathrm{PT}}(r) = \frac{V_0}{\cosh^2(\alpha r)} .
\end{equation}

For the localized bump barriers, we use the compactly supported radial profile
\begin{equation}
  B(q) :=
  \left\{
  \begin{array}{@{}ll@{}}
    \displaystyle
    \exp\left(-\frac{1}{1-q^2}\right), & |q|<1,\\[1ex]
    0, & |q|\geq 1 .
  \end{array}
  \right.
  \label{eq:radial-bump}
\end{equation}
This is a $C^\infty$ bump with compact support in $|q|<1$. We set
\begin{equation}
  T_{\rm off}=t_{\rm on}+\frac{2\pi n}{\omega},
  \qquad
  I(t)=
  \left\{
  \begin{array}{@{}ll@{}}
    1, & t_{\rm on}\le t\le T_{\rm off},\\
    0, & \hbox{otherwise}.
  \end{array}
  \right.
  \label{eq:barrier-time-gate}
\end{equation}
where $n$ is the number of oscillation periods used in the time-dependent cases. The conformal source term used in the evolution is
\begin{equation}
  S_{\rm pot}(\tau,s,\psi)
  =
  -\Omega^{-2}
  V_{\rm barrier} \left(t(\tau,s),r(\tau,s)\right)\psi .
\end{equation}
We write the localized barrier as
\begin{equation}
  V_{\rm barrier}(t,r)
  =
  V_{\rm amp}(t)\,
  B\left(q(t,r)\right),
  \qquad
  q(t,r):=\frac{r-r_{\rm c}(t)}{\delta}.
  \label{eq:barrier-family}
\end{equation}

The three implemented bumps are as follows.
\begin{enumerate}
  \item[(i)] \textbf{Static barrier.}
  The amplitude and center are time-independent:
  \begin{equation}
    V_{\rm amp}(t)=V_0,
    \qquad
    r_{\rm c}(t)=r_0 .
  \end{equation}
  This gives a smooth, spatially compact, time-independent barrier.

  \item[(ii)] \textbf{Pulsating barrier.}
  The center is fixed, while the amplitude oscillates during the active time interval:
  \begin{equation}
    r_{\rm c}(t)=r_0,
    \qquad
    V_{\rm amp}(t)
    =
    I(t)\,\frac{V_0}{2}
    \left[1-\cos \left(\omega(t-t_{\rm on})\right)\right].
    \label{eq:pulsating-amplitude}
  \end{equation}
  Since $T_{\rm off}-t_{\rm on}$ is an integer number of periods, the amplitude vanishes at both endpoints of the active interval.

  \item[(iii)] \textbf{Shaking barrier.}
  The amplitude is fixed, while the center is displaced during the active time interval:
  \begin{equation}
    V_{\rm amp}(t)=V_0,
    \qquad
    r_{\rm c}(t)
    =
    r_0
    +
    I(t)\,\frac{\Delta r}{2}
    \left[1-\cos \left(\omega(t-t_{\rm on})\right)\right].
    \label{eq:shaking-center}
  \end{equation}
  The shaking barrier reduces to the static barrier outside the active interval. During the interval its center moves from $r_0$ to $r_0+\Delta r$ and back.
\end{enumerate}

Unless stated otherwise, we use the incoming Gaussian pulse specified in Section~\ref{sec:eqs_data}. The potential parameters are $V_0=10$ for the scattering runs. For the P\"oschl--Teller model we use $\alpha=3$. For the compact bump-family models we use $r_0=0.5$ and $\delta=0.2$. In the time-dependent cases, $t_{\rm on}=0$, $\omega=5$, and $n=4$.
For the shaking case we take $\Delta r=0.05$.

The potentials above are defined in physical spacetime using standard coordinates. We evaluate them in each patch using the corresponding coordinate maps form Sec.~\ref{sec:geometry}. The P\"oschl--Teller and static bump potentials are active throughout the evolution wherever their spatial support intersects the physical domain. The pulsating barrier is nonzero only for $t_{\rm on}\le t\le T_{\rm off}$ and the shaking barrier differs from the static bump only during that interval. The factor $I(t)$ restricts the pulsating amplitude and the shaking displacement to the active time interval, and all bump barriers are compactly supported in physical radius by $B$.

\begin{figure}[ht]
    \centering
    \includegraphics[width=1\textwidth]{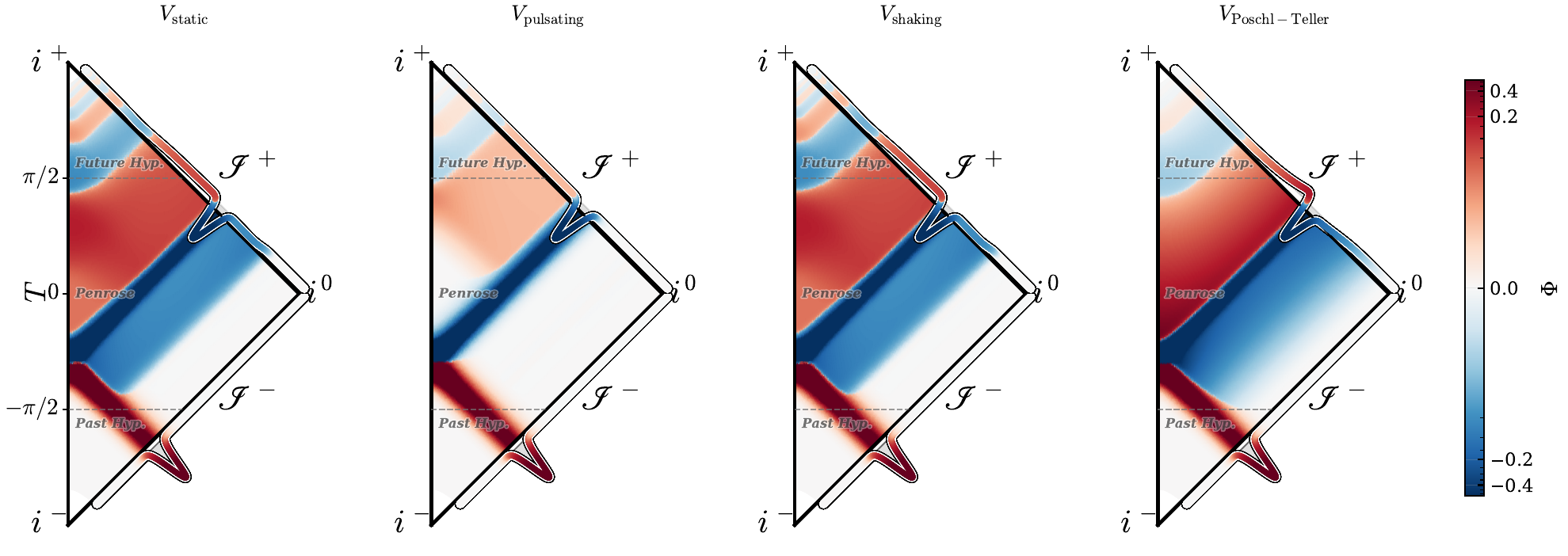}
    \caption{Global scalar field evolution across Penrose diagrams for the four localized potential models described above. The incident Gaussian pulse at past null infinity ($\mathscr{I}^-$) and the corresponding extracted radiation at future null infinity ($\mathscr{I}^+$) are plotted along the null boundaries. The resulting deformations in the asymptotic outgoing profiles show the scattering response of each potential.}
    \label{fig:penrose_potentials}
\end{figure}

The physical effects of these potentials, namely the partial reflection and transmission of the scalar field, are evaluated directly at the asymptotic boundaries. Similarly to the free evolution shown in Fig.~\ref{fig:linear_pulse}, Fig.~\ref{fig:penrose_potentials} displays the global solutions to Eq.~\eref{eq:potential_wave_eqn} as a multipanel grid of Penrose diagrams, with the incoming and outgoing radiation profiles plotted along $\mathscr{I}^-$ and $\mathscr{I}^+$, respectively. By subjecting each configuration to the same initial pulse, the resulting deformations in the outgoing waves encode the scattering dynamics induced by each potential.

\begin{figure}[ht]
    \centering
    \includegraphics[width=0.45\textwidth]{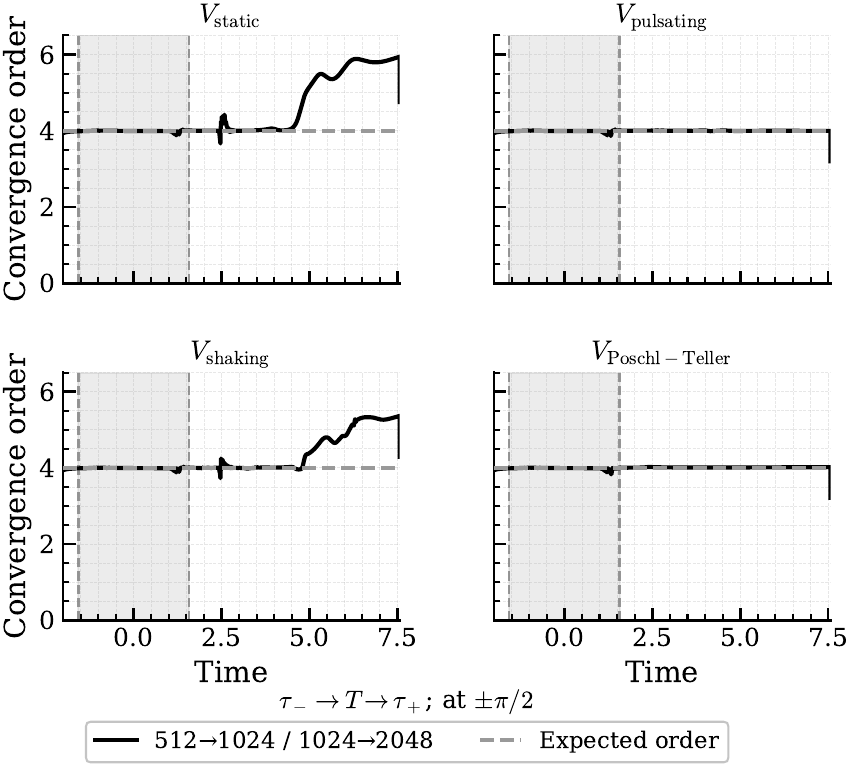}
    \includegraphics[width=0.45\textwidth]{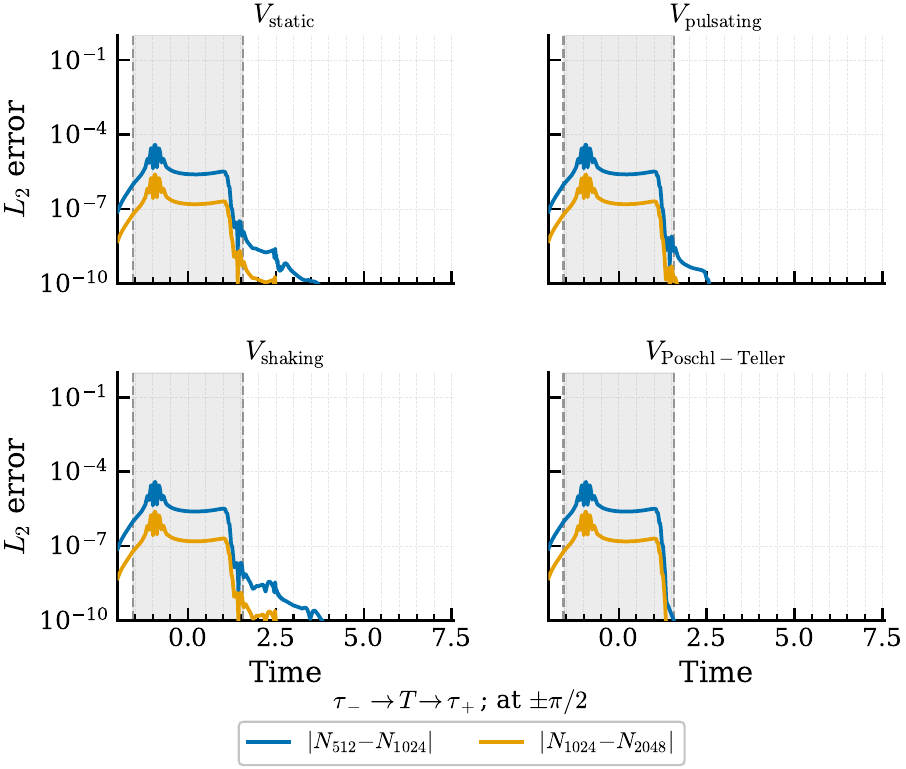}
    
    \caption{\textit{Left Panel:} Self-convergence factor $Q$ for the four localized potentials. \textit{Right Panel:} Pointwise difference between successive resolutions after restriction to a common grid. Norms in the Penrose patch are computed over the active physical interval.}
    \label{fig:convergence_potentials}
\end{figure}

Because no exact solution is available for the potential runs, we perform a self-convergence test. The plotted errors in Fig.~\ref{fig:convergence_potentials} show the difference between successive resolutions after restricting the finer solution to the coarser grid, and the convergence factor $Q$ is computed from the three-resolution comparison in Eq.~\eref{eq:convergence_factor}. In the Penrose patch, all norms are evaluated only over the active physical interval. The scheme shows the expected fourth-order behavior across the multi-patch evolution for every potential profile.

\subsubsection{Quasinormal modes of the P\"oschl--Teller potential}
The quasinormal modes (QNMs) of the P\"oschl--Teller potential are widely studied in the literature both analytically and numerically~\cite{ferrari1984new,Cardona_2017}. The P\"oschl--Teller spectrum is
\begin{equation}
    \omega_n =
    \pm \sqrt{V_0-\frac{\alpha^2}{4}}
    - i\alpha\left(n+\frac{1}{2}\right),
    \qquad n=0,1,2,\ldots .
\end{equation}
In our spherically symmetric 3D framework, the physical wave equation~\eref{eq:waveq_pt} reduces to a 1D wave equation on the half line $r\in[0,\infty)$ after the substitution $w=r\phi$. Regularity at the origin imposes $w(t,0)=0$, selecting the odd-parity sector of the full-line problem. The admissible indices are therefore $n=1,3,5,\ldots$, and the lowest admissible mode corresponds to $n=1$.

In the implementation, the QNM routine analyzes the time series returned by the extraction at future null infinity. The extracted data is $y(T_{\rm g})=\sin R\,\psi|_{\mathscr I^+}$ in the Penrose phase and $y(T_{\rm g})=\sin\rho\,\psi|_{\mathscr I^+}$ in the future hyperboloidal phase, with $T_{\rm g}=T$ in the Penrose patch and $T_{\rm g}=\tau_++\pi/2$ in the future hyperboloidal patch. We discard the early part of this data and keep only samples with $T_{\rm g}>\pi/2+0.05$, so all fit windows used in Table~\ref{tab:qnm_n1} lie in the future hyperboloidal patch. There $u=\tau_+$ at $\mathscr I^+$, and therefore $T_{\rm g}=\tau_++\pi/2$ differs from the physical retarded time only by a constant shift. The fitted frequencies are then the physical P\"oschl--Teller frequencies. The QNM fits quoted in Table~\ref{tab:qnm_n1} were obtained from long P\"oschl--Teller runs at resolution $N=512$. These QNM runs use $A=1$, $\sigma=0.08$, $\tau_0=-0.5$, and $\tau_{-,0}=-3$ for the Gaussian pulse. On a selected ringdown interval we fit
\[
  y(T_{\rm g})
  =
  A_{\rm fit}\exp[-\gamma(T_{\rm g}-T_0)]
  \cos \left[\omega_R(T_{\rm g}-T_0)+\varphi\right],
  \qquad
  \omega_{\rm fit}=\omega_R-i\gamma .
\]
Here $T_0$ is a reference time chosen from the extracted time series, while $A_{\rm fit}$, $\gamma$, $\omega_R$, and $\varphi$ are fitted. The phase $\varphi$ is therefore an independent fit parameter. The analytical frequency is not imposed during the nonlinear least-squares fit. After the fit, the code assigns the overtone index by comparing the fitted damping rate with the P\"oschl--Teller spectrum, using the nearest integer to $|\mathrm{Im}\,\omega_{\rm fit}|/\alpha-1/2$.

\begin{figure} [ht]
    \centering
    \includegraphics[width=0.45\linewidth]{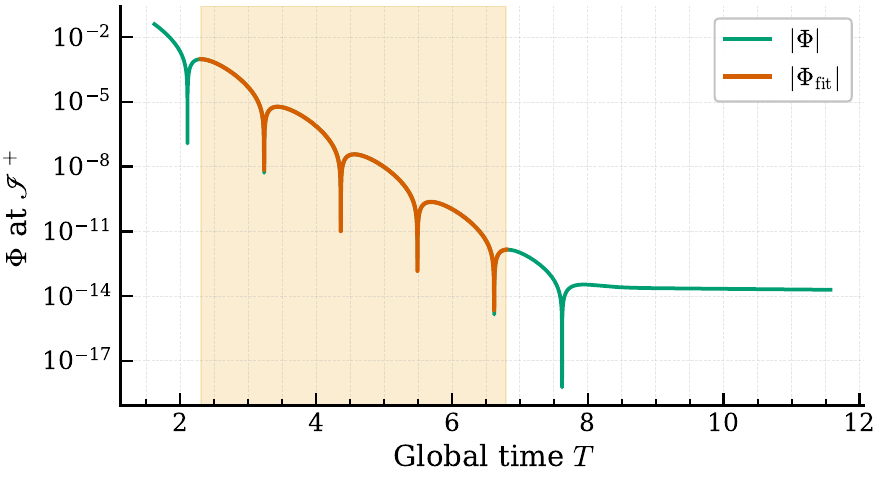}
    \includegraphics[width=0.45\linewidth]{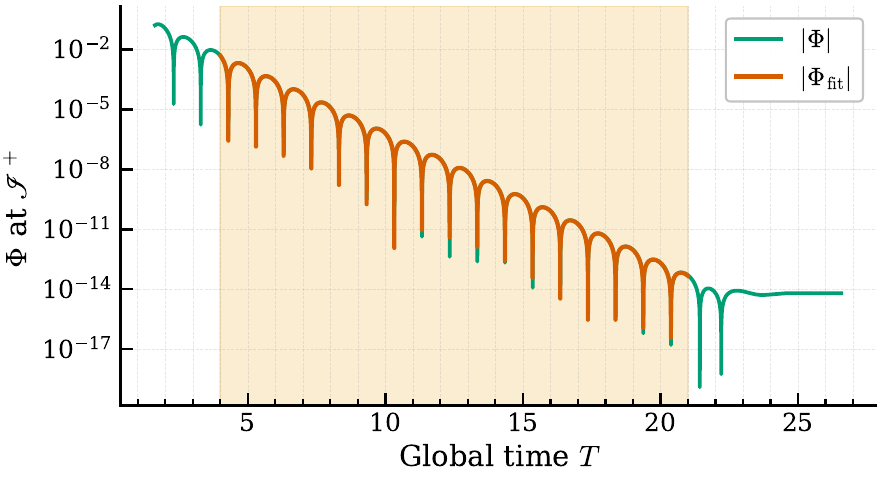}
    \caption{Characteristic quasinormal ringdown of the P\"oschl--Teller potential for two different barrier configurations. \textit{Left panel:} Configuration with $V_0=10$ and $\alpha=3$. The numerical fit gives $\omega_1=2.7837-4.5004i$, differing from the exact analytical value $\omega_1=2.7839-4.5000i$ by $0.005\%$ in the real part and $0.008\%$ in the imaginary part. \textit{Right panel:} Configuration with $V_0=10$ and $\alpha=1$. The numerical fit gives $\omega_1=3.1238-1.5012i$, compared with the analytical value $\omega_1=3.1225-1.5000i$ (relative errors $0.043\%$ and $0.081\%$, respectively).}
    \label{fig:qnm-pt}
\end{figure}

The fit intervals in $T_{\rm g}$ for the four rows of Table~\ref{tab:qnm_n1} are approximately $[2.31,6.80]$, $[4.00,21.00]$, $[3.00,10.00]$, and $[5.00,15.00]$, respectively. These intervals are the actual sampled windows used by the fit. The dominant uncertainty in the fit is the systematic dependence on the chosen ringdown window. Repeating the fits while shifting the endpoints by roughly half an oscillation changes the stable early-ringdown fits by $10^{-5}$--$10^{-2}$ in the fitted frequency components, but later-start windows can be contaminated by the weak tail and may shift the fitted frequency by much larger amounts. The relative errors in Table~\ref{tab:qnm_n1} should therefore be interpreted as deviations for the specified fit windows, not as window-independent error bars. The corresponding time evolution is depicted in Figure~\ref{fig:qnm-pt}, showing characteristic quasinormal ringdown waveforms for two parameter choices as measured directly at future null infinity.

\begin{table}[htbp]
\centering
\caption{Comparison of analytical and numerical quasinormal frequencies ($\omega_1$) for the fundamental odd-parity mode ($n=1$) of the P\"oschl--Teller potential across various barrier parameters.}
\label{tab:qnm_n1}
\begin{tabular*}{\linewidth}{@{\extracolsep{\fill}} l c c c c c @{}}
\toprule
& & & & \multicolumn{2}{c}{Relative Error (\%)} \\
\cmidrule{5-6}
$V_0$ & $\alpha$ & Analytical $\omega_1$ & Numerical $\omega_1$ & Real Part & Imag Part \\
\midrule
10 & 3.0 & $2.7839 - 4.5000i$ & $2.7837 - 4.5004i$ & 0.005 & 0.008 \\
10 & 1.0 & $3.1225 - 1.5000i$ & $3.1238 - 1.5012i$ & 0.043 & 0.081 \\
5 & 2.0 & $2.0000 - 3.0000i$ & $1.9997 - 3.0000i$ & 0.015 & 0.001 \\
5 & 1.0 & $2.1794 - 1.5000i$ & $2.1796 - 1.5001i$ & 0.006 & 0.006 \\
\bottomrule
\end{tabular*}
\end{table}

\subsection{Wave equation with a power nonlinearity}\label{subsec:nonlinear_wave}

To evaluate the framework's capacity to resolve global nonlinear scattering, we turn to the semilinear wave equation in Eq.~\eref{eq:physical-generalV}, characterized by the power-law source $S_{\rm sl}(\phi) = |\phi|^{p-1} \phi$. The nonlinear self-interaction acts as an effective dynamical potential, inducing continuous internal scattering throughout the spacetime volume. This process generates persistent late-time radiation tails, preventing the outgoing field from remaining numerically compactly supported at $\mathscr{I}^+$. Capturing this self-interaction is a challenging test for the code.

\begin{figure}[ht]
    \centering
    \includegraphics[width=0.9\textwidth]{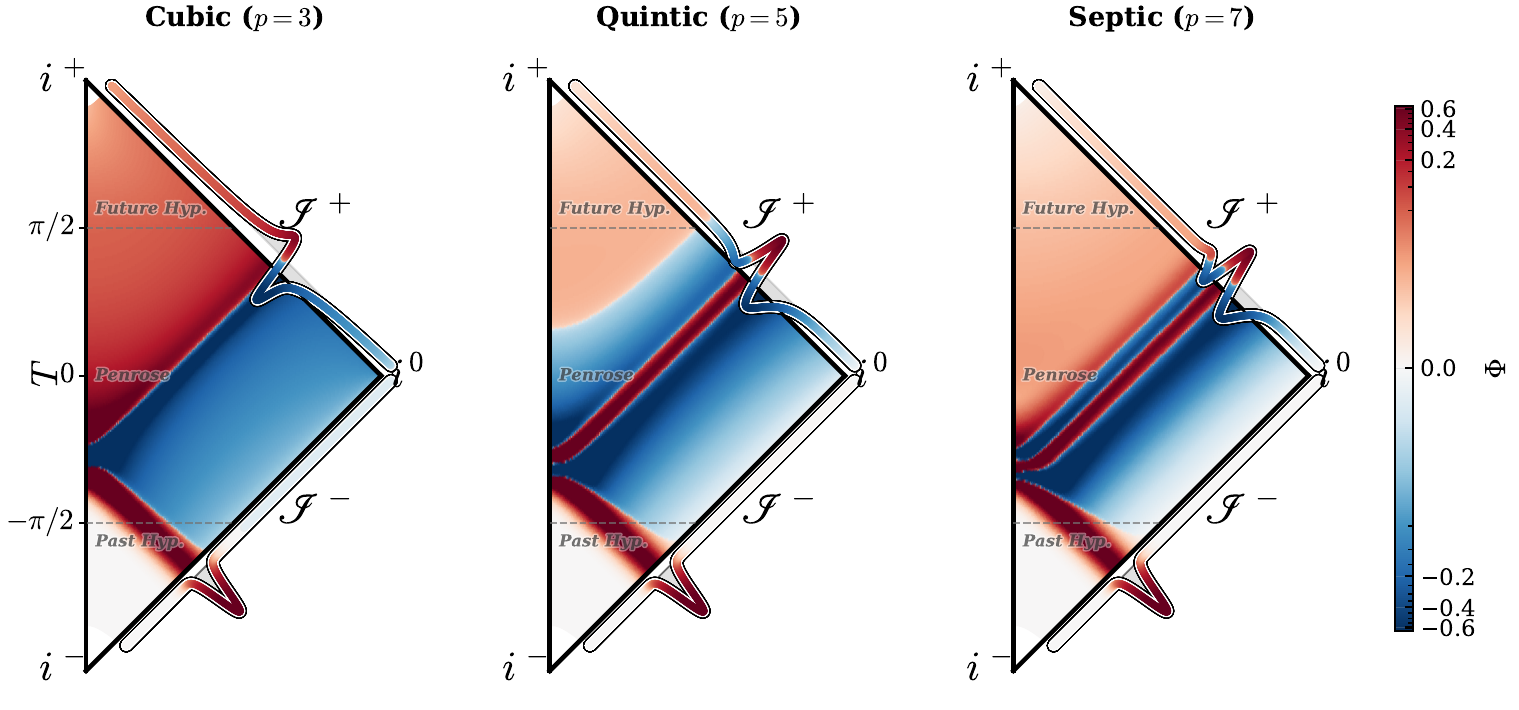}
    \caption{Global scalar field solutions on Penrose diagrams for the cubic ($p=3$), quintic ($p=5$), and septic ($p=7$) semilinear wave equations. }
    \label{fig:higher_order_waveforms}
\end{figure}

We consider the cubic ($p=3$), quintic ($p=5$), and septic ($p=7$) cases. Figure~\ref{fig:higher_order_waveforms} illustrates the global solutions on Penrose diagrams. Unlike the linear free wave, the nonlinear interactions visibly populate the interior of the domain, and the resulting outgoing radiation profiles exhibit long-lived tails at $\mathscr{I}^+$.

\begin{figure}[ht]
    \centering
    \includegraphics[width=1    \textwidth]{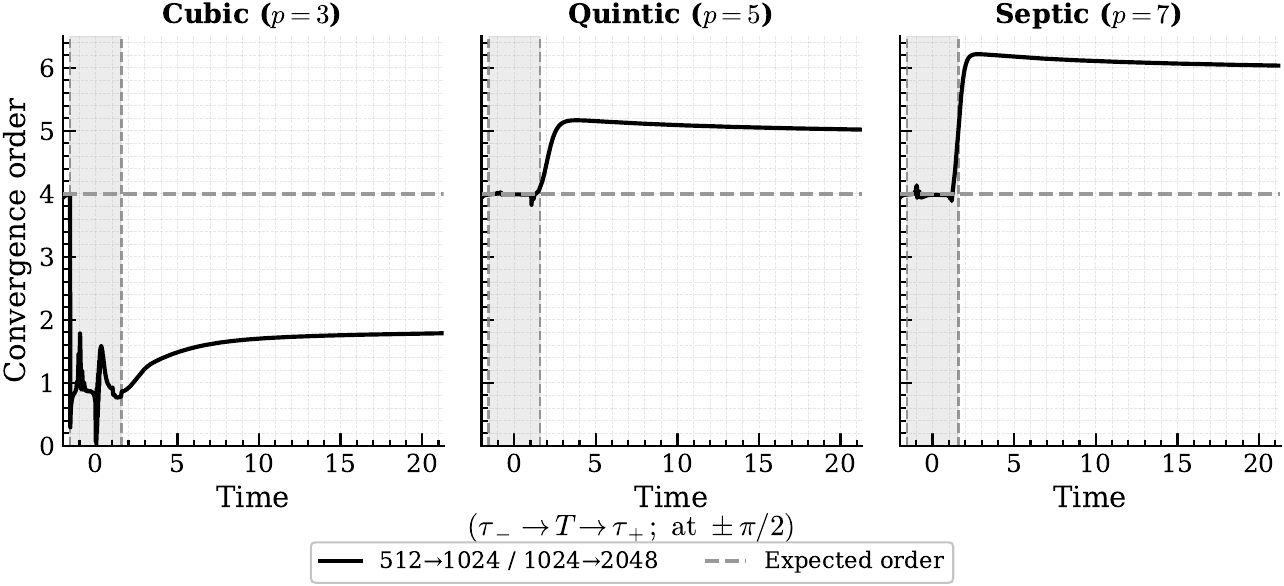}
    \caption{Convergence factor $Q$ for the cubic ($p=3$), quintic ($p=5$), and septic ($p=7$) semilinear wave equations. The quintic and septic cases show approximately fourth-order behavior. The cubic case shows only about first-order convergence, reflecting a limitation of the present finite-difference treatment when the conformally rescaled nonlinear source remains nonzero at the conformal boundary.}
    \label{fig:quintic_convergence}
\end{figure}

The proper resolution of these nonlinearities places stringent demands on the numerical scheme near the conformal boundaries. As shown in Section~\ref{sec:conformal-semilinear}, the conformally rescaled source term scales as $\Omega^{p-3}$. Figure~\ref{fig:quintic_convergence} presents the convergence analysis for these nonlinear models. For the quintic and septic cases, the source term vanishes rapidly at $\mathscr{I}^\pm$, and the scheme maintains approximately fourth-order convergence. The cubic case is different. Since the factor $\Omega^{p-3}$ is unity for $p=3$, the nonlinear source does not vanish at the conformal boundary. Our implementation drops to about first-order convergence in the cubic run. We therefore regard the cubic experiment as a limitation of the current numerical treatment, especially near the compactified interfaces and the point representation of spatial infinity. 

\begin{figure}[ht]
    \centering
    \includegraphics[width=\textwidth]{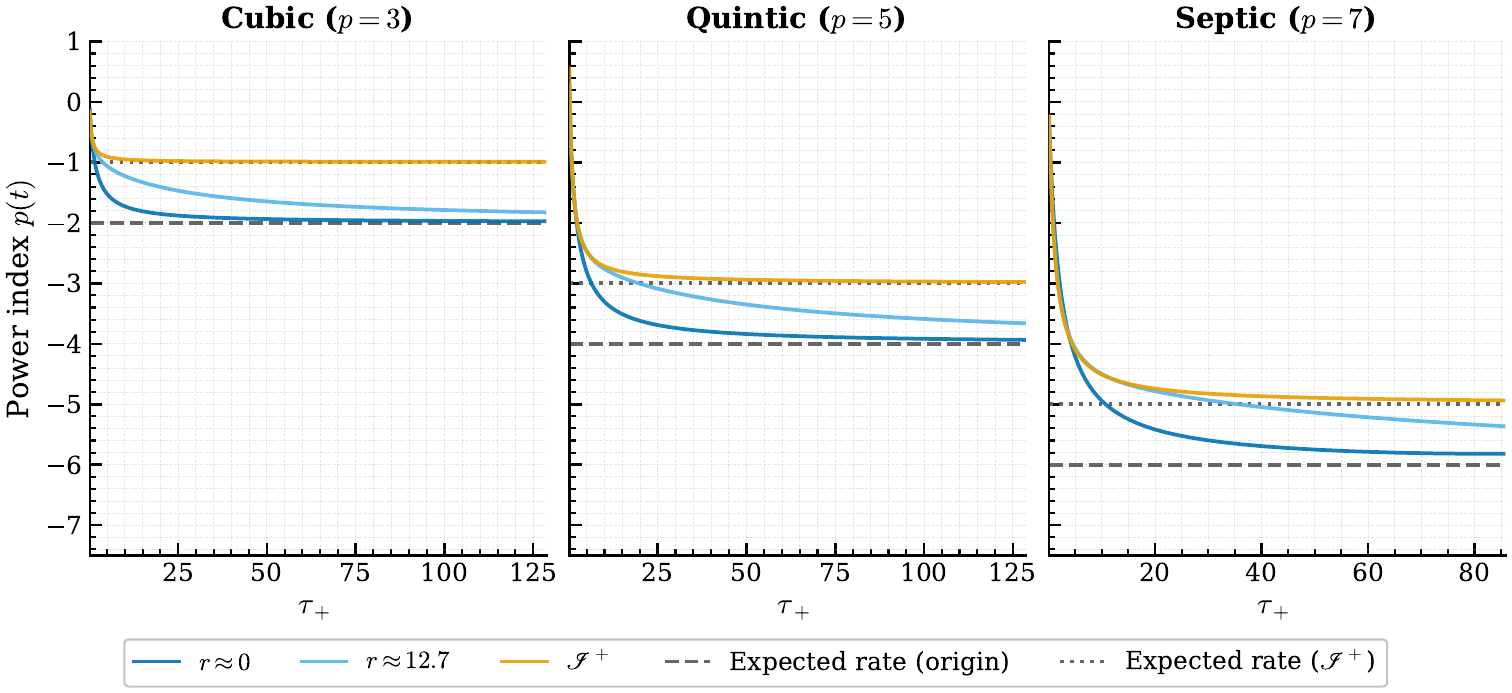}
    \caption{Late-time radiation tails extracted at $\mathscr{I}^+$ for the cubic ($p=3$), quintic ($p=5$), and septic ($p=7$) semilinear wave equations. The agreement with the expected tail rates at late retarded times validates the extraction of nonlinear radiation tails at $\mathscr{I}^+$.}
    \label{fig:nonlinear_tails}
\end{figure}

The accurate extraction of late-time radiation tails is a notoriously difficult numerical problem, often contaminated by spurious boundary reflections in finite-domain methods. Our hybrid approach maps $\mathscr{I}^+$ to a fixed coordinate location, allowing late-time signals to be measured directly at null infinity without artificial outer-boundary reflections. Figure~\ref{fig:nonlinear_tails} demonstrates the late-time polynomial decay of the radiation tails extracted directly at $\mathscr{I}^+$. The expected asymptotic power indices are $-1$ at $\mathscr{I}^+$ and $-2$ at the origin for the cubic equation, $-3$ at $\mathscr{I}^+$ and $-4$ at the origin for the quintic equation, and $-5$ at $\mathscr{I}^+$ and $-6$ at the origin for the septic equation. These rates agree with the known late-time asymptotics for semilinear wave equations~\cite{bizon2009universality,Donninger_2014,Rinne_2025}. Note that, even though the convergence rate is reduced, the code recovers the correct decay rates also for the cubic equation.

\section{Discussion}\label{sec:discussion}

We have presented a global numerical framework for wave scattering from past null infinity to future null infinity in Minkowski spacetime. The construction combines stationary scri-fixed hyperboloidal regions near $\mathscr I^-$ and $\mathscr I^+$ with a central Penrose region resolving a neighborhood of spatial infinity $i^0$. This three-region decomposition is the main geometric framework. The hyperboloidal patches provide time-translation invariant coordinates suitable for long-time evolution at null infinity and the Penrose patch provides a matched conformal bridge between the past and future hyperboloidal foliations.

The Penrose patch mediating between the hyperboloidal domains is not meant to provide the final treatment of spatial infinity for the Einstein equations. In the gravitational problem, the structure of $i^0$ is more subtle. Friedrich's representation of spatial infinity as a cylinder \cite{friedrich1998gravitational} and the associated conformal analysis of regularity provide a more refined framework for the full Einstein problem \cite{Kroon_2003}. Recent fully nonlinear simulations based on Friedrich's generalized conformal field equations have also demonstrated the feasibility of evolving gravitational waves on domains including both past and future null infinity \cite{frauendiener2025fully}. The current work is complementary to these developments. Our goal is to construct a geometric framework that is adapted to long-time simulations with null infinity mapped to the outer boundary of the computational domain. The special simplicity of Minkowski space allowed us to construct a fixed-domain numerical scattering scheme in which the Penrose patch resolves the neighborhood of $i^0$ and matches exactly to the stationary hyperboloidal slices $T=\pm\pi/2\leftrightarrow\tau_\pm=0$. 


In future work, we plan to replace the central Penrose patch by a treatment of spatial infinity better adapted to the conformal Einstein equations, such as the cylinder at spatial infinity based on conformal geodesics, while retaining the stationary scri-fixed hyperboloidal regions for long-time evolution along $\mathscr I^\pm$.  Such a blow-up of $i^0$ may also be beneficial for marginal nonlinear scalar problems, because it separates the directions of approach to spatial infinity instead of representing them by a single grid point. This could provide a more regular numerical setting for the cubic source and is a natural target for improving the first-order behavior observed in Fig.~\ref{fig:quintic_convergence}. Further possible directions include removing the assumption of spherical symmetry and solving the scattering problem on a black-hole spacetime. The black-hole example is particularly interesting because the scattering problem has more channels due to the past and future event horizons. 

In this sense, our work provides a geometric-numerical framwork for formulating scattering problems globally. By separating the long-time hyperboloidal evolution at null infinity from the transition through spatial infinity, the construction provides a numerical method for future treatments of scattering in asymptotically flat spacetimes.

\section*{Acknowledgments}
We thank Alex Vañó-Viñuales and Juan Valiente-Kroon for discussions and comments on the manuscript. ED and AZ are partially supported by the National Science Foundation under Grant No. 2309084. 

\section*{Data and code availability}
The code and data needed to reproduce the figures will be made publicly available upon publication; in the meantime they are available from the corresponding author upon reasonable request.

\section*{References}
\bibliographystyle{iopart-num}
\bibliography{refs}

@misc{compere_asymptotic_2023,
	title = {An asymptotic framework for gravitational scattering},
	publisher = {arXiv},
	author = {Compère, Geoffrey and Gralla, Samuel E. and Wei, Hongji},
	month = nov,
	year = {2023},
	eprint = {2303.17124},
	archive = {arXiv},
	primaryClass = {gr-qc},
}

@article{zenginoglu_hyperboloidal_2008,
	title = {Hyperboloidal foliations and scri-fixing},
	volume = {25},
	doi = {10.1088/0264-9381/25/14/145002},
	number = {14},
	journal = {\pubdoi{10.1088/0264-9381/25/14/145002}{Classical and Quantum Gravity}},
	author = {Zenginoğlu, Anıl},
	month = jul,
	year = {2008},
	pages = {145002},
}

@article{ZENGINOGLU20112286,
title = {Hyperboloidal layers for hyperbolic equations on unbounded domains},
journal = {\pubdoi{10.1016/j.jcp.2010.12.016}{Journal of Computational Physics}},
volume = {230},
number = {6},
pages = {2286-2302},
year = {2011},
doi = {10.1016/j.jcp.2010.12.016},
author = {Anıl Zenginoğlu},
}

@article{penrose_asymptotic_1963,
	title = {Asymptotic {Properties} of {Fields} and {Space}-{Times}},
	volume = {10},
	doi = {10.1103/PhysRevLett.10.66},
	number = {2},
	journal = {\pubdoi{10.1103/PhysRevLett.10.66}{Physical Review Letters}},
	author = {Penrose, Roger},
	month = jan,
	year = {1963},
	pages = {66--68},
}

@article{penrose_zero_1965,
	title = {Zero rest-mass fields including gravitation: asymptotic behaviour},
	volume = {284},
	doi = {10.1098/rspa.1965.0058},
	number = {1397},
	journal = {\pubdoi{10.1098/rspa.1965.0058}{Proceedings of the Royal Society of London. Series A. Mathematical and Physical Sciences}},
	author = {Penrose, Roger},
	month = feb,
	year = {1965},
	pages = {159--203},
}

@article{bizon2009universality,
  title={Universality of global dynamics for the cubic wave equation},
  author={Bizo{\'n}, Piotr and Zengino{\u{g}}lu, An{\i}l},
  journal={Nonlinearity},
  volume={22},
  number={10},
  pages={2473},
  year={2009},
}

@article{zenginoglu_tails_2008,
	title = {A hyperboloidal study of tail decay rates for scalar and {Yang}-{Mills} fields},
	volume = {25},
	doi = {10.1088/0264-9381/25/17/175013},
	number = {17},
	journal = {\pubdoi{10.1088/0264-9381/25/17/175013}{Classical and Quantum Gravity}},
	author = {Zenginoğlu, Anıl},
	year = {2008},
	pages = {175013},
}

@article{Zenginoglu:2024bzs,
    author = "Zengino\u{g}lu, An\i{}l",
    title = "{Hyperbolic times in Minkowski space}",
    eprint = "2404.01528",
    archive = "arXiv",
    primaryClass = "gr-qc",
    doi = "10.1119/5.0214271",
    journal = {\pubdoi{10.1119/5.0214271}{Am. J. Phys.}},
    volume = "92",
    pages = "965--974",
    year = "2024"
}

@misc{Kadar:2025xmo,
  title={Scattering, Polyhomogeneity and Asymptotics for Quasilinear Wave Equations From Past to Future Null Infinity},
  author={Kadar, Istvan and Kehrberger, Lionor},
  year={2025},
  eprint={2501.09814},
  archive={arXiv}
}

@article{Lovelace_2009,
   title={Reducing spurious gravitational radiation in binary-black-hole simulations by using conformally curved initial data},
   volume={26},
   doi={10.1088/0264-9381/26/11/114002},
   number={11},
   journal={\pubdoi{10.1088/0264-9381/26/11/114002}{Classical and Quantum Gravity}},
   author={Lovelace, Geoffrey},
   year={2009},
   month=may, pages={114002} }

@article{Kroon_2003,
doi = {10.1088/0264-9381/20/5/102},
year = {2003},
volume = {20},
number = {5},
pages = {L53},
author = {Juan Antonio Valiente Kroon},
title = {Early radiative properties of the developments of time-symmetric conformally flat initial data},
journal = {\pubdoi{10.1088/0264-9381/20/5/102}{Classical and Quantum Gravity}},
}

@article{Pfeiffer_2003,
   title={Extrinsic curvature and the {Einstein} constraints},
   volume={67},
   doi={10.1103/physrevd.67.044022},
   number={4},
   journal={\pubdoi{10.1103/physrevd.67.044022}{Physical Review D}},
   author={Pfeiffer, Harald P. and York, James W.},
   year={2003},
   month=feb }

@article{duarte2026numerical,
  title={A numerical approach to particle creation in accelerating toy models},
  author={Duarte-Baptista, Pedro and Va{\~n}{\'o}-Vi{\~n}uales, Alex and del R{\'\i}o, Adri{\'a}n},
  journal={Classical and Quantum Gravity},
  volume={43},
  number={4},
  pages={045009},
  year={2026},
}

@article{Panosso_Macedo_2024,
   title={Hyperboloidal approach for static spherically symmetric spacetimes: a didactical introduction and applications in black-hole physics},
   volume={382},
   doi={10.1098/rsta.2023.0046},
   number={2267},
   journal={\pubdoi{10.1098/rsta.2023.0046}{Philosophical Transactions of the Royal Society A: Mathematical, Physical and Engineering Sciences}},
   author={Panosso Macedo, Rodrigo},
   year={2024},
   month=jan }

@book{Kroon_2023, place={Cambridge}, title={Conformal Methods in General Relativity}, publisher={Cambridge University Press}, author={Kroon, Juan A. Valiente}, year={2023}}

@article{valiente2024d,
  title={The {d'Alembert} solution in hyperboloidal foliations},
  author={Valiente Kroon, Juan A and Gomes Da Silva, Lidia J},
  journal={General Relativity and Gravitation},
  volume={56},
  number={7},
  pages={85},
  year={2024},
}

@article{friedlander_radiation_1980,
  author  = {Friedlander, F. G.},
  title   = {Radiation fields and hyperbolic scattering theory},
  journal = {\pubdoi{10.1017/S0305004100057819}{Mathematical Proceedings of the Cambridge Philosophical Society}},
  volume  = {88},
  number  = {3},
  pages   = {483--515},
  year    = {1980},
  doi     = {10.1017/S0305004100057819},
}

@article{baez_global_1990,
  author  = {Baez, John C. and Segal, Irving E. and Zhou, Zheng-Fang},
  title   = {The global {Goursat} problem and scattering for nonlinear wave equations},
  journal = {\pubdoi{10.1016/0022-1236(90)90128-8}{Journal of Functional Analysis}},
  volume  = {93},
  number  = {2},
  pages   = {239--269},
  year    = {1990},
  doi     = {10.1016/0022-1236(90)90128-8},
}

@article{mason_nicolas_2004,
  author  = {Mason, Lionel J. and Nicolas, Jean-Philippe},
  title   = {Conformal scattering and the {Goursat} problem},
  journal = {\pubdoi{10.1142/S0219891604000123}{Journal of Hyperbolic Differential Equations}},
  volume  = {1},
  number  = {2},
  pages   = {197--233},
  year    = {2004},
  doi     = {10.1142/S0219891604000123},
}

@article{lindblad_schlue_2023,
  author  = {Lindblad, Hans and Schlue, Volker},
  title   = {Scattering from infinity for semilinear wave equations satisfying the null condition or the weak null condition},
  journal = {\pubdoi{10.1142/S0219891623500066}{Journal of Hyperbolic Differential Equations}},
  volume  = {20},
  number  = {1},
  pages   = {155--218},
  year    = {2023},
  doi     = {10.1142/S0219891623500066},
}

@article{lindblad_schlue_2025,
  author  = {Lindblad, Hans and Schlue, Volker},
  title   = {Scattering for wave equations with sources close to the light cone and prescribed radiation fields},
  journal = {\pubdoi{10.2140/paa.2025.7.865}{Pure and Applied Analysis}},
  volume  = {7},
  number  = {4},
  pages   = {865--926},
  year    = {2025},
  doi     = {10.2140/paa.2025.7.865},
}

@article{gasperin_gautam_hilditch_vano_2020,
  author  = {Gasperin, Edgar and Gautam, Shalabh and Hilditch, David and {Va{\~n}{\'o}-Vi{\~n}uales}, Alex},
  title   = {The hyperboloidal numerical evolution of a good-bad-ugly wave equation},
  journal = {\pubdoi{10.1088/1361-6382/ab5f21}{Classical and Quantum Gravity}},
  volume  = {37},
  number  = {3},
  pages   = {035006},
  year    = {2020},
  doi     = {10.1088/1361-6382/ab5f21},
}

@article{gautam_vano_hilditch_bose_2021,
  author  = {Gautam, Shalabh and {Va{\~n}{\'o}-Vi{\~n}uales}, Alex and Hilditch, David and Bose, Sukanta},
  title   = {Summation by parts and truncation error matching on hyperboloidal slices},
  journal = {\pubdoi{10.1103/PhysRevD.103.084045}{Physical Review D}},
  volume  = {103},
  number  = {8},
  pages   = {084045},
  year    = {2021},
  doi     = {10.1103/PhysRevD.103.084045},
}

@article{peterson_gautam_vano_hilditch_2024,
  author  = {Peterson, Christian and Gautam, Shalabh and {Va{\~n}{\'o}-Vi{\~n}uales}, Alex and Hilditch, David},
  title   = {Spherical evolution of the generalized harmonic gauge formulation of general relativity on compactified hyperboloidal slices},
  journal = {\pubdoi{10.1103/PhysRevD.110.124033}{Physical Review D}},
  volume  = {110},
  number  = {12},
  pages   = {124033},
  year    = {2024},
  doi     = {10.1103/PhysRevD.110.124033},
}

@article{Cardona_2017,
   title={Quasinormal modes of generalized {Pöschl}–{Teller} potentials},
   volume={34},
   doi={10.1088/1361-6382/aa9428},
   number={24},
   journal={\pubdoi{10.1088/1361-6382/aa9428}{Classical and Quantum Gravity}},
   author={Cardona, A F and Molina, C},
   year={2017},
   month=Nov, pages={245002} }

@article{Rinne_2025,
   title={A hyperboloidal method for numerical simulations of multidimensional nonlinear wave equations: nonlinear tails},
   volume={38},
   doi={10.1088/1361-6544/ae153e},
   number={10},
   journal={\pubdoi{10.1088/1361-6544/ae153e}{Nonlinearity}},
   author={Rinne, Oliver},
   year={2025},
   month=Oct, pages={105026} }

@article{Donninger_2014,
   title={Nondispersive decay for the cubic wave equation},
   volume={7},
   doi={10.2140/apde.2014.7.461},
   number={2},
   journal={\pubdoi{10.2140/apde.2014.7.461}{Analysis \& PDE}},
   author={Donninger, Roland and Zenginoğlu, Anil},
   year={2014},
   month=May, pages={461–495} }

@article{frauendiener2025fully,
  title={Fully nonlinear gravitational wave simulations from past to future null infinity},
  author={Frauendiener, J{\"o}rg and Stevens, Chris and Thwala, Sebenele},
  journal={Physical Review Letters},
  volume={134},
  number={16},
  pages={161401},
  year={2025},
}

@misc{marajh_taujanskas_valientekroon_2025,
  title={Controlled regularity at future null infinity from past asymptotic initial data: the wave equation},
  author={Marajh, Jordan and Taujanskas, Grigalius and Kroon, Juan A Valiente},
  year={2025},
  eprint={2508.04690},
  archive={arXiv}
}

@book{carroll2019spacetime,
  title={Spacetime and geometry},
  author={Carroll, Sean M},
  year={2019},
  publisher={Cambridge University Press}
}

@article{friedrich1998gravitational,
  title={Gravitational fields near space-like and null infinity},
  author={Friedrich, Helmut},
  journal={Journal of Geometry and Physics},
  volume={24},
  number={2},
  pages={83--163},
  year={1998},
}

@article{ferrari1984new,
  title={New approach to the quasinormal modes of a black hole},
  author={Ferrari, Valeria and Mashhoon, Bahram},
  journal={Physical Review D},
  volume={30},
  number={2},
  pages={295},
  year={1984},
}

@misc{zenginouglu2026penrose,
  title={From {Penrose} to {Melrose}: computing scattering amplitudes at infinity for unbounded media},
  author={Zengino{\u{g}}lu, An{\i}l},
  year={2026},
  eprint={2601.04167},
  archive={arXiv}
}

@article{frauendiener2004conformal,
  title={Conformal infinity},
  author={Frauendiener, J{\"o}rg},
  journal={Living Reviews in Relativity},
  volume={7},
  number={1},
  pages={1},
  year={2004},
}

@article{friedrich1983cauchy,
  title={Cauchy problems for the conformal vacuum field equations in general relativity},
  author={Friedrich, Helmut},
  journal={Communications in Mathematical Physics},
  volume={91},
  number={4},
  pages={445--472},
  year={1983},
}

@article{friedrich1981asymptotic,
  title={The asymptotic characteristic initial value problem for {Einstein}’s vacuum field equations as an initial value problem for a first-order quasilinear symmetric hyperbolic system},
  author={Friedrich, Helmut},
  journal={Proceedings of the Royal Society of London. A. Mathematical and Physical Sciences},
  volume={378},
  number={1774},
  pages={401--421},
  year={1981},
}

@article{bishop_rezzolla_2016,
  author        = {Bishop, Nigel T. and Rezzolla, Luciano},
  title         = {Extraction of Gravitational Waves in Numerical Relativity},
  journal       = {\pubdoi{10.1007/s41114-016-0001-9}{Living Reviews in Relativity}},
  volume        = {19},
  number        = {1},
  pages         = {2},
  year          = {2016},
  doi           = {10.1007/s41114-016-0001-9},
  eprint        = {1606.02532},
  archive = {arXiv},
  primaryClass  = {gr-qc}
}

@article{bishop_gomez_lehner_maharaj_winicour_1996,
  author        = {Bishop, Nigel T. and G{\'o}mez, Roberto and Lehner, Luis
                   and Maharaj, Manoj and Winicour, Jeffrey},
  title         = {Cauchy-characteristic extraction in numerical relativity},
  journal       = {\pubdoi{10.1103/PhysRevD.54.6153}{Physical Review D}},
  volume        = {54},
  pages         = {6153--6165},
  year          = {1996},
  doi           = {10.1103/PhysRevD.54.6153},
  eprint        = {gr-qc/9705033},
  archive = {arXiv}
}

@article{moxon2023spectre,
  title={{SpECTRE} {C}auchy-characteristic evolution system for rapid, precise waveform extraction},
  author={Moxon, Jordan and Scheel, Mark A and Teukolsky, Saul A and Deppe, Nils and Vu, Nils and H{\'e}bert, Francois and Kidder, Lawrence E and Throwe, William},
  journal={Physical Review D},
  volume={107},
  number={6},
  pages={064013},
  year={2023},
}

@article{bernuzzi2025perturbative,
  title={Perturbative hyperboloidal extraction of gravitational waves in 3+ 1 numerical relativity},
  author={Bernuzzi, Sebastiano and Fontbut{\'e}, Joan and Albanesi, Simone and Zengino{\u{g}}lu, An{\i}l},
  journal={Physical Review D},
  volume={112},
  number={8},
  pages={084036},
  year={2025},
}

@article{Alvares:2025pbi,
    author = "{\'A}lvares, Jo{\~a}o D. and Va{\~n}{\'o}-Vi{\~n}uales, Alex",
    title = "{Charged scalar field at future null infinity via nonlinear hyperboloidal evolution}",
    eprint = "2506.15311",
    archivePrefix = "arXiv",
    primaryClass = "gr-qc",
    doi = "10.1103/w94x-d6j8",
    journal = "Phys. Rev. D",
    volume = "112",
    number = "10",
    pages = "104053",
    year = "2025",
    note = "[Erratum: Phys.Rev.D 113, 049902 (2026)]"
}

@article{Vano-Vinuales:2024tat,
    author = "Va{\~n}{\'o}-Vi{\~n}uales, Alex and Valente, Tiago",
    title = "{Height-function-based 4D reference metrics for hyperboloidal evolution}",
    eprint = "2408.08952",
    archivePrefix = "arXiv",
    primaryClass = "gr-qc",
    doi = "10.1007/s10714-024-03323-8",
    journal = "Gen. Rel. Grav.",
    volume = "56",
    number = "11",
    pages = "135",
    year = "2024"
}

@article{Vano-Vinuales:2014koa,
    author = "Va{\~n}{\'o}-Vi{\~n}uales, Alex and Husa, Sascha and Hilditch, David",
    title = "{Spherical symmetry as a test case for unconstrained hyperboloidal evolution}",
    eprint = "1412.3827",
    archivePrefix = "arXiv",
    primaryClass = "gr-qc",
    doi = "10.1088/0264-9381/32/17/175010",
    journal = "Class. Quant. Grav.",
    volume = "32",
    number = "17",
    pages = "175010",
    year = "2015"
}

@article{Zenginoglu:2008pw,
    author = "Zenginoglu, Anil",
    title = "{Hyperboloidal evolution with the Einstein equations}",
    eprint = "0808.0810",
    archivePrefix = "arXiv",
    primaryClass = "gr-qc",
    reportNumber = "AEI-2008-050",
    doi = "10.1088/0264-9381/25/19/195025",
    journal = "Class. Quant. Grav.",
    volume = "25",
    pages = "195025",
    year = "2008"
}

@article{bizon2010saddle,
  title={Saddle-point dynamics of a Yang--Mills field on the exterior Schwarzschild spacetime},
  author={Bizo{\'n}, Piotr and Rostworowski, Andrzej and Zengino{\u{g}}lu, An{\i}l},
  journal={Classical and Quantum Gravity},
  volume={27},
  number={17},
  pages={175003},
  year={2010}
}

@article{ripley2021numerical,
  title={Numerical computation of second-order vacuum perturbations of Kerr black holes},
  author={Ripley, Justin L and Loutrel, Nicholas and Giorgi, Elena and Pretorius, Frans},
  journal={Physical Review D},
  volume={103},
  number={10},
  pages={104018},
  year={2021},
  publisher={APS}
}

@article{bourg2025quadratic,
  title={Quadratic quasinormal modes at null infinity on a Schwarzschild spacetime},
  author={Bourg, Patrick and Macedo, Rodrigo Panosso and Spiers, Andrew and Leather, Benjamin and Bonga, B{\'e}atrice and Pound, Adam},
  journal={Physical Review D},
  volume={112},
  number={4},
  pages={044049},
  year={2025},
  publisher={APS}
}

\appendix

\section{The free solution and its conformal representations}
\label{sec:free-test-solution}

In this section we construct an explicit non-trivial spherically symmetric solution of the standard flat wave equation
\begin{equation}
  \Box_\eta \phi = 0,
  \label{eq:flat-wave}
\end{equation}
and rewrite it as a solution of the corresponding conformal wave equations in (i) hyperboloidal coordinates $(\tau_\pm,\rho)$ and (ii) Penrose coordinates $(T,R)$. Throughout, we assume spherical symmetry with $\Delta_\sigma \psi = 0$.

\subsection{A smooth spherically symmetric Minkowski solution}\label{app:A1}
For a spherically symmetric field $\phi=\phi(t,r)$ in $3+1$ Minkowski,
\begin{equation}
  \Box_\eta \phi
  = -\partial_t^2\phi + \partial_r^2\phi + \frac{2}{r}\partial_r\phi
  = \frac{1}{r}\left(-\partial_t^2+\partial_r^2\right)\left(r\phi\right).
  \label{eq:radial-wave-factorisation}
\end{equation}
Hence, if $w:=r\phi$ solves the $1+1$ wave equation $( -\partial_t^2+\partial_r^2 )w=0$, then $\phi=w/r$ solves (\ref{eq:flat-wave}) for $r>0$. A convenient family of \emph{regular} solutions at $r=0$ is obtained by taking
\begin{equation}
  w(t,r)=F(t-r)-F(t+r),
  \qquad
  \phi(t,r)=\frac{F(t-r)-F(t+r)}{r},
  \label{eq:phi-general}
\end{equation}
with $F\in C^\infty(\mathbb R)$. Regularity at the center follows from
\begin{equation}
  \phi(t,0):=\lim_{r\to 0}\frac{F(t-r)-F(t+r)}{r}=-2F'(t).
  \label{eq:phi-regular-origin}
\end{equation}
To obtain an explicit non-trivial pulse with strong decay in the null directions, we fix
\begin{equation}
  F(s)=e^{-s^2}.
  \label{eq:F-gaussian}
\end{equation}
Then
\begin{equation}
  \phi(t,r)=\frac{e^{-(t-r)^2}-e^{-(t+r)^2}}{r}
  \label{eq:phi-explicit}
\end{equation}
The solution is smooth at $r=0$ with $\phi(t,0)=4t\,e^{-t^2}$. In terms of the null coordinates $u=t-r$ and $v=t+r$, the associated reduced field is simply
\begin{equation}
  w(t,r)=r\phi(t,r)=e^{-u^2}-e^{-v^2}.
  \label{eq:w-null}
\end{equation}
The raw limits of the reduced field are therefore
\begin{equation}
  \lim_{r\to\infty,\; v\ \mathrm{fixed}} w(t,r)
  =
  -F(v),
  \qquad
  \lim_{r\to\infty,\; u\ \mathrm{fixed}} w(t,r)
  =
  F(u).
  \label{eq:raw-free-radiation-limits}
\end{equation}
This sign difference is a consequence of regularity at the center: the regular radial solution is represented by the odd extension $w(t,r)=F(t-r)-F(t+r)$. With the radiation-field normalization \eqref{eq:radiation-field-normalization}, the scattering data are
\begin{equation}
  \mathcal R_-(v)=-F(v),
  \qquad
  \mathcal R_+(u)=-F(u).
\end{equation}
Hence the free scattering map is the identity after identifying the natural parameters on the two null infinities, $\mathcal R_+(s)=\mathcal R_-(s)$. If instead one compared the raw traces of $w$ at $\mathscr I^-$ and $\mathscr I^+$, the corresponding free map would differ by an overall minus sign.

\subsubsection*{Hyperboloidal representation}
We use the hyperboloidal compactification (\ref{eq:past_hyperboloidal}) and (\ref{eq:future_hyperboloidal}) to the Minkowski solution (\ref{eq:phi-explicit}) and $\psi_\pm:=\Omega_H^{-1}\phi=\phi/\cos\rho$. Note that
\begin{equation}
  r\,\Omega_H = \tan\rho\,\cos\rho=\sin\rho,
\end{equation}
so that
\begin{equation}
  \psi_\pm(\tau_\pm,\rho)=\frac{w(t(\tau_\pm,\rho),r(\rho))}{\sin\rho}.
  \label{eq:psi-hyperboloidal-via-w}
\end{equation}
Substituting $w=e^{-u^2}-e^{-v^2}$ from (\ref{eq:w-null}) and using the characteristic expressions (\ref{eq:uv_future}) and (\ref{eq:uv_past}) yields
\begin{equation}
  \psi_+(\tau_+,\rho)
  =
  \frac{
    e^{-\left(\tau_+ + \frac{\cos\rho}{1+\sin\rho}\right)^2}
    -
    e^{-\left(\tau_+ + \frac{1+\sin\rho}{\cos\rho}\right)^2}
  }{\sin\rho},
  \label{eq:psi-plus-explicit}
\end{equation}
and
\begin{equation}
  \psi_-(\tau_-,\rho)
  =
  \frac{
    e^{- \left(\tau_- - \frac{1+\sin\rho}{\cos\rho}\right)^2}
    -
    e^{-\left( \tau_- - \frac{\cos\rho}{1+\sin\rho}\right)^2}
  }{\sin\rho}.
  \label{eq:psi-minus-explicit}
\end{equation}
These functions are smooth at $\rho=0$ because the numerator vanishes linearly with $\sin\rho$; the extension to $\rho=\pi/2$ encodes the radiation fields. Incoming data at $\scri^-$ reads
\begin{equation}
  \lim_{\rho\to\pi/2}\sin\rho\,\psi_-(\tau_-,\rho)= -e^{-\tau_-^2}.
 \end{equation}
Outgoing data at $\scri^+$ reads
\begin{equation}
  \lim_{\rho\to\pi/2}\sin\rho\,\psi_+(\tau_+,\rho)=e^{-\tau_+^2}.
\end{equation}

\subsubsection*{Penrose representation}
Define $\psi:=\Omega_P^{-1}\phi=\phi/(\cos T+\cos R)$. Using $r\Omega_P=\sin R$ we again have
\begin{equation}
  \psi(T,R)=\frac{w(t(T,R),r(T,R))}{\sin R}.
  \label{eq:psi-penrose-via-w}
\end{equation}
Substituting (\ref{eq:w-null}) yields the explicit Penrose-coordinate conformal field
\begin{equation}
  \psi(T,R) =
  \frac{
    e^{-\tan^2\left(\frac{T-R}{2}\right)}
    -
    e^{-\tan^2\left(\frac{T+R}{2}\right)}
  }{\sin R}.
  \label{eq:psi-penrose-explicit}
\end{equation}
This expression is smooth at $R=0$ by the same cancellation mechanism as at $\rho=0$.

\subsection{A Gaussian incoming pulse prescribed at past null infinity}
\label{sec:gaussian-incoming}

For numerical tests across the interface at $\tau_-=0$ (equivalently $T=-\pi/2$), it is convenient to prescribe an incoming pulse on $\mathscr I^-$ that is already negligible by the time $\tau_-$ approaches $0$. We fix parameters $\tau_0<0$, $\sigma>0$, and $A\in\mathbb R$, and define the advanced-time Gaussian profile
\begin{equation}
  G(s) := A \exp\left(-\frac{(s-\tau_0)^2}{2\sigma^2}\right).
  \label{eq:G-gaussian}
\end{equation}
This choice corresponds to taking $F=-G$ in the free solution of \ref{app:A1}. Equivalently, the reduced field is
\begin{equation}
  w(t,r)=G(v)-G(u),
\end{equation}
so that the incoming field at $\mathscr I^-$ is $G$ and the raw outgoing field at $\mathscr I^+$ is $-G$. With the radiation-field normalization \eqref{eq:radiation-field-normalization}, both the incoming and outgoing radiation fields are equal to $G$. Choosing $|\tau_0|/\sigma\gg 1$ makes the incident data exponentially small at $\tau_-=0$ (e.g.\ $|\tau_0|/\sigma\gtrsim 9$ gives $|G(0)|\lesssim 10^{-18}|A|$). In this sense, the Gaussian pulse is numerically compactly supported once its amplitude falls below machine precision. The past hyperboloidal solution reads
\begin{equation}
  \psi_-(\tau_-,\rho)
  =
  \frac{
    G\left(\tau_- - \frac{\cos\rho}{1+\sin\rho}\right)
    -
    G\left(\tau_- - \frac{1+\sin\rho}{\cos\rho}\right)
  }{\sin\rho},
  \label{eq:psi-minus-gaussian}
\end{equation}
The future hyperboloidal solution reads
\begin{equation}
  \psi_+(\tau_+,\rho)
  =
  \frac{
    G\left(\tau_+ + \frac{1+\sin\rho}{\cos\rho}\right)
    -
    G\left(\tau_+ + \frac{\cos\rho}{1+\sin\rho}\right)
  }{\sin\rho}.
  \label{eq:psi-plus-gaussian}
\end{equation}

In the same way, we obtain the explicit Penrose-coordinate solution
\begin{equation}
  \psi(T,R)
  =
  \frac{
    G\left(\tan\left(\frac{T+R}{2}\right)\right)
    -
    G\left(\tan\left(\frac{T-R}{2}\right)\right)
  }{\sin R}.
  \label{eq:psi-penrose-gaussian}
\end{equation}

At $\mathscr I^-$, i.e.\ $\rho\to\pi/2$ in the past chart, we obtain
\begin{equation}
  \lim_{\rho\to\pi/2}\psi_-(\tau_-,\rho)
  =
  G(\tau_-),
\end{equation}
and hence
\begin{equation}
  \mathcal R_-(\tau_-)=G(\tau_-).
\end{equation}
At $\mathscr I^+$, i.e.\ $\rho\to\pi/2$ in the future chart, the raw trace is
\begin{equation}
  \lim_{\rho\to\pi/2}\psi_+(\tau_+,\rho)
  =
  -G(\tau_+).
\end{equation}
Therefore
\begin{equation}
  \mathcal R_+(\tau_+)
  =
  -\lim_{\rho\to\pi/2}\psi_+(\tau_+,\rho)
  =
  G(\tau_+).
\end{equation}
Thus the Gaussian profile prescribed as incoming radiation at $\mathscr I^-$ is recovered as the outgoing radiation field at $\mathscr I^+$ in the free evolution.

\end{document}